\documentclass[notitlepage,english,aps,tightenlines,floatfix,12pt,citeautoscript]{revtex4}

\usepackage{graphicx}
\usepackage{dcolumn}
\usepackage{bm}
\usepackage{color}
\usepackage[colorlinks=true,urlcolor=blue,citecolor=blue, linkcolor = blue]{hyperref}
\usepackage{amssymb,amsmath}
\usepackage[normalem]{ulem}
\usepackage[italicdiff]{physics}

\newcommand{\qql}{\textquotedblleft}
\newcommand{\qqr}{\textquotedblright}

\newcommand{\dg}{\dagger}
\newcommand{\ham}{\hat{\mathcal{H}}}
\newcommand{\vam}{\hat{\mathcal{V}}}

\begin{document}
	
	\title{Superconductivity, generalized random phase approximation and linear scaling methods}
	
	\author{Sebastiano Peotta}
	
	\affiliation{Department of Applied Physics, Aalto University, FI-00076 Aalto, Finland}
	
	
	\begin{abstract}
		The superfluid weight is an important observable of superconducting materials since it is related to the London penetration depth of the Meissner effect. It can be computed from the change in the grand potential (or free energy) in response to twisted boundary conditions in a torus geometry. 
		Here we review the Bardeen-Cooper-Schrieffer mean-field theory emphasizing its origin as a variational approximation for the grand potential. The variational parameters are the effective fields that enter in the mean-field Hamiltonian, namely the Hartree-Fock potential and the pairing potential. The superfluid weight is usually computed by ignoring the dependence of the effective fields on the twisted boundary conditions. However, it has been pointed out in recent works that this can lead to unphysical results, particularly in the case of lattice models with flat bands. As a first result, we show that taking into account the dependence of the effective fields on the twisted boundary conditions leads in fact to the generalized random phase approximation. 
		Our second result is providing the mean-field grand potential as an explicit function of the one-particle density matrix. This allows us to derive the expression for the superfluid weight within the generalized random phase approximation in a transparent manner. Moreover, reformulating mean-field theory as a well-posed minimization problem in terms of the one-particle density matrix is a first step towards the application  to superconducting systems of the linear scaling methods developed in the context of electronic structure theory. 
	\end{abstract}
	
	\maketitle
	

	\section{Introduction}
	
	The most striking manifestation of superconductivity is the vanishing of electrical resistence below the critical temperature. Quantum mechanics predicts that in a perfectly crystalline solid the resistance is zero, at least at zero temperature where phononic excitations are frozen out, however this ideal situation is never realized in practice since impurities and defects in the crystal structure are present in any material. Impurities and crystalline defects, hereafter collectively called disorder, are responsible for the finite resistence in the zero temperature limit in normal metals. On the other hand, superconducting materials are perfect conductors below the critical temperature as if disorder vanished altogether. Nevertheless, disorder has a significant effect on many of the observable properties of superconductors and when increased above a certain threshold it drives a transition to the normal state, which can be metallic or even insulating~\cite{Sadovskii1997,Imry1981,Maekawa1984,Ma1986,Ghosal1998,Ghosal2001}. The interplay between superconductivity and disorder is a vast research subject and many questions remain open~\cite{Zhao2019,Sacepe2020,Lau2022}. 
	
	The phenomenon of perfect conductivity and the electromagnetic response of superconducting materials are well described by combining Maxwell's equations with the constitutive relation $\vb{J}= -D_{\rm s}\vb{A}$ introduced by Heinz and Fritz London~\cite{London1935}. This is a peculiar relation since a gauge invariant quantity, the current density $\vb{J}$, is proportional to the vector potential $\vb{A}$, which is not  gauge invariant.  This is not wrong since the London constitutive relation is valid only in a specific gauge, the London gauge~\cite{London1935}.
	The coefficient $D_{\rm s}$ is known as the superfluid weight and it  has been argued that a nonzero superfluid weight is the defining property of a superconductor~\cite{Scalapino1992,Scalapino1993}. Indeed, beside dissipationless transport, the phenomenon that defines superconductivity is the Meissner effect, the expulsion of the magnetic field from the bulk of a superconductor. In the superconducting state, the magnetic field can only penetrate up to a characteristic length scale $\lambda_{\rm L}$ away from the surface of the sample. This so-called London penetration depth is simply the superfluid weight expressed in different units since the two quantities are related by $\lambda_{\rm L} = (\mu_0D_{\rm s})^{-1/2}$, with $\mu_0$ the magnetic constant. Moreover, the superfluid weight determines the critical temperature in a two-dimensional superfluid according to the Nelson-Kosterliz relation~\cite{Nelson1977}.
	
	As explained below in Section~\ref{sec:superfluid_weight}, the superfluid weight is obtained as a specific limit of the current-current response functions~\cite{Scalapino1992,Scalapino1993}. An alternative and equivalent definition is as the change of the grand potential (or free energy) under a twist of the boundary conditions~\cite{Fisher1973,Scalapino1993,Tovmasyan2018} (see Eq.~\eqref{eq:d2F_dA2} below). Twisted boundary conditions can be introduced by means of a constant vector potential $\vb{A}$ in a torus geometry, which cannot be gauged away since it amounts to nonzero magnetic fluxes through the holes of the torus (see Section~\ref{sec:basics}).
	
	In general, one has to resort to approximations since computing exactly either the current-current response function or the grand potential of a quantum many-body system is an intractable task. The standard approximation used to describe superconducting materials is mean-field theory, known also as  Bardeen-Cooper-Schrieffer (BCS) theory~\cite{Bardeen1957,Bardeen1957a,Schrieffer1964,Tinkham2004}. Within BCS theory, the quartic interaction term in the many-body Hamiltonian is replaced by a quadratic one (see Eq.~\eqref{eq:H_BdG} below), which depends on effective fields $\Gamma_{i,j},\,\Delta_{i,j}$ that have to be computed self-consistently. In particular, the pairing term in the mean-field Hamiltonian has the form $\frac{1}{2}\sum_{i,j}\big(\Delta_{i,j} \hat{c}_{i}^\dg \hat{c}_{j}^\dg + \Delta_{i,j}^*\hat{c}_{j}\hat{c}_{i}\big)$ and is essential for superconducting systems since the pairing potential $\Delta_{i,j}$ plays the role of the order parameter of the superconducting transition. 
	
	Typically, the mean-field approximation for $D_{\rm s}$ is obtained by replacing the many-body Hamiltonian $\mathcal{\hat H}$ with the mean-field Hamiltonian $\mathcal{\hat H}_{\rm m.f.}$ in the current-current response function, which can then be evaluated using Wick's theorem~\cite{Scalapino1993,Liang2017a}. This procedure is equivalent to taking the partial derivatives of the mean-field grand potential $\Omega_{\rm m.f.}(\vb{A},\Delta,\Gamma)$ with respect to the constant vector potential $\vb{A}$ associated to twisted boundary conditions, more precisely $D_{{\rm s}, lm} \propto \pdv*{\Omega_{\rm m.f.}(\vb{A},\Delta,\Gamma)}{A_l}{A_m}|_{\vb{A} = \vb{0}}$ (in anisotropic systems the superfluid weight is a tensor in general). However, it is known that the current-current response function computed using $\mathcal{\hat{H}}_{\rm m.f.}$ breaks gauge invariance~\cite{Scalapino1993} and, as noted in recent works~\cite{Huhtinen2022,Chan2022a}, it can give clearly unphysical results in the case of multiorbital lattice models, in particular when flat bands are present in the band structure. 
	
	The problem is that the effective fields $\Gamma_{i,j},\,\Delta_{i,j}$ are obtained by minimizing the mean-field grand potential for each separate value of $\vb{A}$, which means that they are themselves functions of $\vb{A}$, but this dependence is usually neglected~\cite{Scalapino1993,Peotta2015,Liang2017a}. In Ref.~\onlinecite{Huhtinen2022}, it was shown how the full second derivative of $\Omega_{\rm m.f.}\qty\big(\vb{A},\Gamma(\vb{A}),\Delta(\vb{A}))$ with respect to $\vb{A}$ can be evaluated in terms of the solution of the mean-field problem for $\vb{A} = \vb{0}$ only. Here we go a step further than Ref.~\onlinecite{Huhtinen2022} since the first main result of this work is to show that computing the superfluid weight as the full second derivative of the mean-field grand potential (see Eq.~\eqref{eq:D_mf} below) is in fact equivalent to another well-known gauge invariant approximation, the generalized random phase approximation for superconducting systems introduced by Anderson~\cite{Anderson1958} and Rickayzen~\cite{Rickayzen1959}. The expression for the superfluid weight in the generalized random phase approximation is given below in Eq.~\eqref{eq:main_Ds_GRPA} and is a result not contained in previous works, in particular in Refs.~\onlinecite{Huhtinen2022,Chan2022a}, where the importance of taking into account the $\vb{A}$-dependence of the effective fields has been pointed out.
	
	In order to derive the expression of the superfluid weight in the generalized random phase approximation~\eqref{eq:main_Ds_GRPA}, we start by reviewing the BCS mean-field theory of superconductivity  in Section~\ref{sec:mean-field}, using as a starting point the variational principle at finite temperature. At finite temperature the variational principle is encoded in the Bogoliubov inequality (Eq.~\eqref{eq:Bogoliubov_ineq} below), from which it follows that the mean-field grand potential is an upper bound for the exact grand potential $\Omega(\vb{A})\leq \Omega_{\rm m.f.}\qty\big(\vb{A},\Gamma,\Delta)$~\eqref{eq:Bogoliubov_second}. Thus the mean-field solution, that is the choice of the effective fields $\Gamma_{i,j},\,\Delta_{i,j}$ minimizing $\Omega_{\rm m.f.}$, gives the best possible upper bound on the exact grand potential. The fact that mean-field theory is a variational approximation for the grand potential makes it clear that the latter is a better starting point for computing the superfluid weight rather than the current-current response function. This is important to justify the results of Section~\ref{sec:superfluid_weight}. Whereas the results of Section~\ref{sec:mean-field} are standard, namely the self-consistency equations of BCS theory~\eqref{eq:BCS_self_cons_1}-\eqref{eq:BCS_self_cons_3},
	their derivation is more rigorous than the one usually found in standard textbooks~\cite{Schrieffer1964,Parravicini2013} and provides a better starting point for the derivation of the original results of this work, contained in Sections~\ref{sec:mermin_functional} and~\ref{sec:superfluid_weight}. 
	
	To derive the self-consistency equations of BCS theory in Section~\ref{sec:mean-field}, we find it convenient to perform a Legendre transform and switch from the effective fields $\Gamma_{i,j},\,\Delta_{i,j},\,\Delta_{i,j}^*$ to the expectation values $\expval*{\hat{c}_{i}^\dg\hat{c}_{j}},\,\expval*{\hat{c}_{i}^\dg \hat{c}_{j}^\dg},\,\expval*{\hat{c}_{j}\hat{c}_{i}}$ as the variational parameters. These are the matrix elements of the one-particle density matrix $P$. In particular the anomalous (not number-conserving) expectation values, $\langle \hat {c}_i \hat{c}_j\rangle$ and $\langle \hat{c}_i^\dg \hat{c}_j^\dg \rangle$ are essential to describe superconducting systems. In Section~\ref{sec:mermin_functional}, we go a step further and obtain the mean-field grand potential as an explicit  function of the one-particle density matrix (see Eqs.~\eqref{eq:Omega_0_bar_P_final}-\eqref{eq:Omega_LWBK} below), which is a natural generalization of the functional used by Mermin to prove the extension of the Hohenberg-Kohn theorems to finite temperature~\cite{Mermin1965}. Moreover, we identify precisely the class of one-particle density matrices within which the variational optimization is performed. In this way the mean-field problem is recast as a mathematically well-posed minimization problem in terms of the one-particle density matrix. This is the second original result of this work. 
 	
 	It turns out that the derivation of the expression for the superfluid weight in the generalized random phase approximation~\eqref{eq:main_Ds_GRPA} is particularly transparent if the starting point is the mean-field potential $\Omega_{\rm m.f.}\qty\big(\vb{A},P(\vb{A}))$ as a function of the one-particle density matrix, where $P(\vb{A})$ is the density matrix which minimizes $\Omega_{\rm m.f.}(\vb{A},P)$ for a given value of the constant vector potential $\vb{A}$. This is the approach used in Section~\ref{sec:superfluid_weight}. Equivalently, one can start from the grand potential expressed in the terms of the effective fields $\Omega_{\rm m.f.}\qty\big(\vb{A},\Gamma(\vb{A}),\Delta(\vb{A}))$, as done in Ref.~\onlinecite{Huhtinen2022}, but then the derivation of~\eqref{eq:main_Ds_GRPA} becomes rather cumbersome, which  is probably the reason why the connection with the random phase approximation has not been unveiled in previous works. This illustrates the advantage of reformulating mean-field theory in terms of the one-particle density matrix.
 	
 	The original results presented in Section~\ref{sec:mermin_functional} are potentially important for another reason, namely they are a first step towards the application of linear scaling methods to superconducting systems. Linear scaling means that the computational resources (time and memory) required for the numerical solution of a problem scale linearly $\sim O(N)$ with the number of constituents $N$, for instance the lattice sites in a tight-binding model. The numerical solution of the mean-field problem generally scales as $O(N^3)$ since it requires the full diagonalization of the mean-field Hamiltonian. This scaling severely limits the accessible system size~\cite{Lau2022}. This is a problem when studying disordered superconductors since it is generally necessary to consider rather large systems. The need to simulate large systems has been further exacerbated with the discovery of superconductivity in moir\'e heterostructures~\cite{Cao2018,Torma2022}, such as magic-angle twisted bilayer graphene, which can contain thousands of atoms in a moir\'e unit cell. The na\"ive application of mean-field theory in this case is computationally  very demanding and one has to introduce further approximations, such as restricting to a limited number of moir\'e minibands and/or renormalization procedures, which have been shown to lead to problems when evaluating important observables such as the superfluid weight~\cite{Hu2019,Julku2020,Xie2020}. Moreover, the simulation of moir\'e heterostructures in the presence of disorder is currently out of reach.
	 
	 In the field of electronic structure theory, many different approaches have been developed to improve the scaling of density functional theory and empirical tight-binding calculations. An important class of methods, known as $O(N)$- or linear scaling methods~\cite{Goedecker1999,Bowler2012}, takes advantage of the fact that the density matrix is ranged, i.e., it decays with distance
	 \begin{equation}
	 	\label{eq:density_matrix_decay}
	 	P(\vb{r},\vb{r}') \to 0,\qquad |\vb{r}-\vb{r}'|\to \infty\,.
	 \end{equation}
	 The basic approximation behind linear scaling methods is to neglect the matrix elements  for large separation, for instance $|\vb{r}-\vb{r}'|> R$, and thus represent $P$ as a sparse matrix. This approximation is justified only if the decay in~\eqref{eq:density_matrix_decay} is sufficiently fast. Quantifying the exact decay behavior of the density matrix is a nontrivial problem that has received a lot of attention and definite results have been obtained. For instance, in insulators one has an exponential decay $P(\vb{r},\vb{r}') \sim e^{-\gamma|\vb{r}-\vb{r}'|}$, where the decay rate $\gamma$ can be related to the energy gap~\cite{Goedecker1999}.  If the range $R$ is fixed, the number of nonzero matrix elements of $P$ scales linearly with the system size and the computational effort required to find the optimal value of these parameters scales linearly as well. From a numerical point of view, for achieving linear scaling it is necessary to use standard sparse matrix methods for matrix multiplication and for the solution of linear systems of equations. 
	 
	 Linear scaling methods can take advantage of massively parallelized high performance computing and have been successfully applied to systems with millions of atoms, which are relevant for material science and biology. On the other hand, linear scaling methods have not been employed so far in the field of superconductivity. To this end, it is essential to reformulate standard BCS mean-field theory as a minimization problem in which the independent variable is the one-particle  density matrix $P$, as done in Section~\ref{sec:mermin_functional}. The essential new ingredient is the inclusion in the density matrix of the anomalous (not number-conserving) expectation values, $\langle \hat {c}_i \hat{c}_j\rangle$ and $\langle \hat{c}_i^\dg \hat{c}_j^\dg \rangle$, that are the trademark of BCS theory. In Section~\ref{sec:mermin_functional} we also discuss why it should be possible to directly apply to superconductive systems the linear scaling methods that have been developed so far.
	 
	 The structure of this work is as follows. In Section~\ref{sec:basics}, we introduce the class of lattice models we are interested in, we define the current density operator and explain how to represent twisted boundary conditions by means of a constant vector potential. The concept of gauge invariance is discussed as well. In Section~\ref{sec:mean-field}, we derive BCS mean-field theory starting from the Bogoliubov inequality to emphasize its origin as a variational approximation for the grand potential. In Section~\ref{sec:mermin_functional}, we express the mean-field grand potential as an explicit function of the one-particle density matrix: the main result is given by~\eqref{eq:Omega_0_bar_P_final}-\eqref{eq:Omega_LWBK}. We conclude the section by arguing that, based on our results, the application of linear scaling method to superconducting systems should be rather straightforward. In Section~\ref{sec:superfluid_weight}, we compute the superfluid weight by taking the full second derivative of the mean-field grand potential with respect to the constant vector potential and show that this leads to a gauge invariant result~\eqref{eq:main_Ds_GRPA}, which we identify with the generalized random phase approximation. This is compared in Appendix~\ref{app:equivalence} to the standard mean-field formula for the superfluid weight found in the literature. In Section~\ref{sec:conclusion}, we summarize our results and discuss possible directions for future work.
 	 
	 \section{Basic definitions}
	 \label{sec:basics}
	 
	Here we consider a generic tight-binding Hamiltonian 
    with a static interaction term. More precisely, the non-interacting part of the Hamiltonian takes the form
	\begin{equation}
		\label{eq:Ham_free}
	\mathcal{\hat H}_{\rm free} = \sum_{i,j} \hat{c}_i^\dg K_{i,j} \hat{c}_j\,,	
	\end{equation}   
	where $\hat{c}_i$ and $\hat{c}_i^\dg$ are the usual fermionic fields operators and $K$ is the hopping matrix collecting all the hopping matrix elements ($K_{i,j}$ for $i\neq j$) and the on-site potential ($K_{i,i}$). Here  $i,j$ are collective indices labeling all the degrees of freedom in the model (spatial, orbital, spin, \dots). We adopt the convention that symbols with a hat, such as $\mathcal{\hat H}_{\rm free}$ and $\hat{c}_i$,  denote operators in the many-body Fock space, while operators on the single-particle Hilbert space, such as the hopping matrix $K$, do not have hats.
	
	The interaction term is quadratic in the occupation number operators $\hat{n}_i = \hat{c}_i^\dg\hat{c}_i$
	\begin{equation}
		\label{eq:Ham_int}
		\mathcal{\hat H}_{\rm int} = \frac{1}{2}\sum_{i,j}V_{i,j}\hat{n}_i\hat{n}_j\,.
	\end{equation}
	The factor 1/2 in the above equation eliminates double-counting since we require the interaction potential to be symmetric $V_{i,j} = V_{j,i}$ and we also set $V_{i,i} = 0$. One may consider other types of interaction terms, and we leave to the interested reader the task of extending our arguments to more general settings. The full many-body Hamiltonian is then the sum of~\eqref{eq:Ham_free} and~\eqref{eq:Ham_int}
	\begin{equation}
		\label{eq:Ham_tot}
	\mathcal{\hat H} = \mathcal{\hat H}_{\rm free} + \mathcal{\hat H}_{\rm int}\,.	
	\end{equation}

\begin{figure}
		\includegraphics[scale=1]{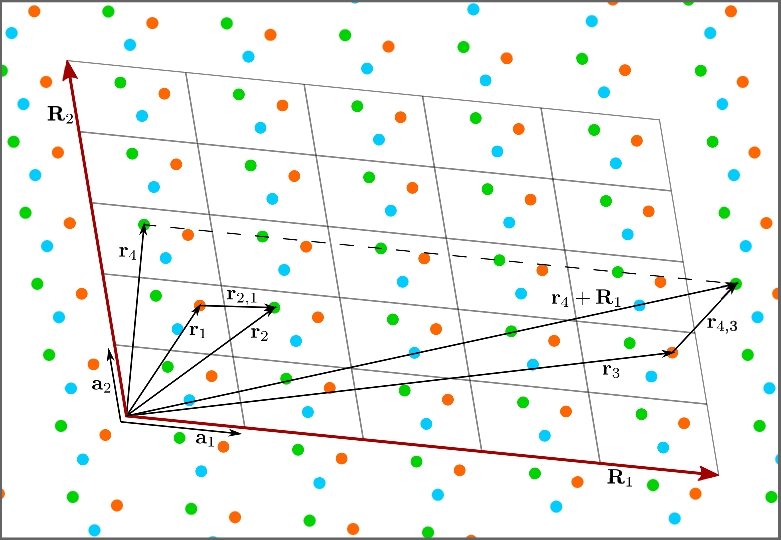}
		\caption{\label{fig:lattice} A periodic lattice $L = \{\vb{r}_j\}$, is a discrete collection of points $\vb{r}_j$ invariant under translations by vectors in $\Gamma = \mathrm{Span}_{\mathbb{Z}}(\vb{a}_1,\vb{a}_2)$. $\Gamma$ is the Bravais lattice generated by the fundamental vectors $\vb{a}_1$ and $\vb{a}_2$. For simplicity only two-dimensional lattices are considered. The lattice shown in the figure is composed of three different sublattices represented as dots of different colors. To introduce periodic boundary conditions, two noncollinear Bravais lattice vectors $\vb{R}_1,\,\vb{R}_2\in \Gamma$ ($\vb{R}_{i} = 5\vb{a}_i$ in the figure) are selected. Then, two points $\vb{r}_i,\vb{r}_j\in L$ are identified if $\vb{r}_i-\vb{r}_j = m_1\vb{R}_1+m_2\vb{R}_2$, for some integers $m_1$ and $m_2$ (for instance the two points joined by a dashed line in the figure). Denote by $[\vb{r}] = \big\{ \vb{r} + m_1\vb{R}_1 + m_2\vb{R}_2\, \big|\, m_1,\,m_2 \in \mathbb{Z}\big\}$ the equivalence classes obtained by this identification. In the case of periodic boundary conditions, an equivalence class $[\vb{r}_i]$  is attached to each site $i$   and the distance between the sites is defined as $d(i,j) = \mathrm{min}\big\{|\vb{r}|\,\big|\,\vb{r} \in [\vb{r}_i-\vb{r}_j]\big\}$. This is known as the flat torus distance. The displacement vector going from site $j$ to site $i$ is the vector $\vb{r}_{i,j} \in [\vb{r}_i-\vb{r}_j]$ with the property that $d(i,j) = |\vb{r}_{i,j}|$, if this vector is unique. If there are multiple vectors with this property, the displacement vector is not defined. If $d(i,j)\ll |\vb{R}_1|,\,|\vb{R}_2|$ the displacement vector $\vb{r}_{i,j}$ is always defined. As shown in the figure, the displacement vector from site 1 to site 2 is $\vb{r}_{2,1} = \vb{r}_2-\vb{r}_1$, while the one from site $3$ to site $4$ is $\vb{r}_{4,3} = \vb{r}_4-\vb{r}_3+\vb{R}_1\neq \vb{r}_4-\vb{r}_3$.}
	\end{figure}

	Linear scaling methods take advantage of the locality of the hopping matrix~\eqref{eq:Ham_free} and of the density matrix~\eqref{eq:density_matrix_decay}. Locality entails some notion of distance between different degrees of freedom, which is introduced by embedding the tight-binding model in space.  In the case of infinitely extended systems or a finite system with open boundary conditions, a position vector $\vb{r}_j$ is assigned to the degree of freedom labeled by $j$ and the distance $d(i,j) = |\vb{r}_{i,j}|$ is the length of the displacement vector $\vb{r}_{i,j} = \vb{r}_i-\vb{r}_j$. The position vectors take values in a periodic lattice $L$, which can be a composite (non-Bravais) lattice as shown in Figure~\ref{fig:lattice}. In the following we find convenient to use the word \qql site\qqr as an alternative to \qql degrees of freedom,\qqr but it is understood that site $i$ and site $j$ may by assigned to the same lattice point $\vb{r}_i = \vb{r}_j$. Physically, this is the case when the two sites are in fact different electronic orbitals on the same atom or the two different spin states of the same atomic orbital. For simplicity, the geometric arrangement of the sites is given by a periodic lattice. However the Hamiltonian is not necessary periodic, in particular disorder is usually introduced as a random perturbation of the matrix $K$.

	In the case of periodic boundary conditions, the points of the lattice $L$ are identified if they differ by integer linear combinations of  two Bravais lattice vectors $\vb{R}_1$ and $\vb{R}_2$ (we consider only two-dimensional systems here), which define the size and shape of the finite system, called also the the \qql supercell.\qqr An equivalence class $[\vb{r}_j]$ obtained under this identification, that is a point on a torus, is attached to each site $j$. In the case of periodic boundary conditions, the distance $d(i,j)$ and the displacement vectors $\vb{r}_{i,j}$  are defined in the caption of Figure~\ref{fig:lattice}. We provide a precise definition of the displacement vectors in Figure~\ref{fig:lattice} since they are essential to introduce twisted boundary condition and the current density operator, see below.
	
	In realistic cases, the hopping matrix elements between atomic orbitals decrease exponentially with distance,  namely  $K_{i,j} \sim e^{-\alpha d(i,j)}$ for some $\alpha > 0$, therefore it is practical to consider only hopping matrices with finite range $R$, that is $K_{i,j} = 0$ if $d(i,j) > R$. It is assumed in the following that the hopping matrix has a range of the order of a few lattice constants and, in the case of periodic boundary conditions, the range is much smaller than the linear size of the supercell $R\ll |\vb{R}_1|, |\vb{R}_2|$. 
	
	Using the continuity equation for the occupation number operators $\hat{n}_j$,
	one identifies the microscopic current operator
	\begin{equation}
		\hat{J}_{i,j} = -\hat{J}_{j,i} =  \frac{1}{\hbar}\left(-i\hat{c}_i^\dagger K_{i,j}\hat{c}_j + \mathrm{H.c.}\right)\,,
	\end{equation} 	
	whose expectation value gives the particle current flowing from site $j$ to site $i$. We set $\hbar = 1$ from now on. One is mainly interested in the Fourier transform of the current density, which in the long-wavelength limit ($\vb{q} \to \vb{0}$) reads 
	\begin{equation}
	\label{eq:Jq}
	\hat{\vb{J}}(\vb{q}) = \frac{1}{2\mathcal{A}}\sum_{i,j}\vb{r}_{i,j}  \hat{J}_{i,j} e^{-i\vb{q}\cdot\frac{\vb{r}_i+\vb{r}_j}{2}}\,.
	\end{equation}
	For periodic boundary conditions, $\mathcal{A} = |\vb{R}_1\times \vb{R}_2|$ is the area of the supercell (for dimension $d=2$) and the wavevector $\vb{q}$ belongs to the reciprocal lattice of the Bravais lattice generated by $\vb{R}_1, \vb{R}_2$, meaning that $\vb{q}\cdot \vb{R}_i = 2\pi m_i$ with $m_i$ an integer.

	The conservation of current is a consequence of the gauge symmetry of electromagnetism, which in a discrete lattice model consists in the invariance under unitary transformations $\mathcal{\hat{U}}_{\rm g}$ acting on the fields  operators as 
	\begin{equation}
		\label{eq:gauge_transf}
		\mathcal{\hat{U}}_{\rm g} \hat{c}_{j} \mathcal{\hat{U}}_{\rm g}^\dg = e^{i\theta_j}\hat{c}_{j}\,,
	\end{equation}  
	where $\theta_j$ are arbitrary site-dependent phases. Some operators change form under gauge transformations ($\hat{\vb{J}}(\vb{q})$, $\mathcal{\hat{H}}_{\rm free}$), while others do not ($\hat{n}_j$, $\mathcal{\hat{H}}_{\rm int}$). Observable quantities (expectation values) are always gauge invariant. Consider the following modified noninteracting Hamiltonian
	\begin{equation}
		\label{eq:Ham_free_A}
		\mathcal{\hat H}_{\rm free}(\vb{A}) = \sum_{i,j} \hat{c}_i^\dg K_{i,j}(\vb{A}) \hat{c}_j = \sum_{i,j} \hat{c}_i^\dg K_{i,j}e^{iq\vb{A}\cdot \vb{r}_{i,j}} \hat{c}_j\,.
	\end{equation}
	The Peierls phase $e^{iq\vb{A}\cdot \vb{r}_{i,j}}$ accounts for the presence of a constant vector potential $\vb{A}$. The particle charge is set to $q = 1$ from now on. In the case of an infinitely extended system or a system with open boundary conditions, a constant vector potential is a pure gauge, which does not affect the observable properties in any way. Indeed, since $\vb{r}_{i,j} = \vb{r}_{i} -\vb{r}_{j}$, the Hamiltonian~\eqref{eq:Ham_free_A} is obtained by performing a gauge transformation of the form $\mathcal{\hat{U}}_{\rm g}(\vb{A}) \hat{c}_{j} \mathcal{\hat{U}}_{\rm g}^\dg(\vb{A}) = e^{-i\vb{A}\cdot\vb{r}_j}\hat{c}_{j}$, namely $\mathcal{\hat{U}}_{\rm g}(\vb{A}) \mathcal{\hat H}_{\rm free}(\vb{A}=\vb{0}) \mathcal{\hat{U}}_{\rm g}^\dg(\vb{A}) = \mathcal{\hat H}_{\rm free}(\vb{A})$. On the other hand, there is in general no analogous gauge transformation in the case of periodic boundary conditions. The reason is that even a constant vector potential amounts to nonzero magnetic fluxes through the holes of the torus in which the lattice is embedded: these fluxes are given by $\Phi_{i} = \vb{A}\cdot \vb{R}_{i}$ with $i = 1,2$. The magnetic flux in lattice models is defined up to integer multiples of the magnetic flux quantum $\phi_0 = 2\pi \hbar/q$ ($=2\pi$ in our units) and is invariant under gauge transformations. It follows that two Hamiltonians $\mathcal{\hat H}_{\rm free}(\vb{A})$ and $\mathcal{\hat H}_{\rm free}(\vb{A}')$ can be mapped into each other by a gauge transformation if and only if the respective magnetic fluxes differ by an integer multiple of the magnetic flux quantum: $\Phi_i-\Phi_i' = (\vb{A}-\vb{A}')\cdot \vb{R}_i = 2\pi m_i$ with $i = 1,2$ and $m_i$ arbitrary integers. Using a gauge transformation, it can be shown that a constant vector potential is equivalent to twisted boundary conditions parametrized by the two magnetic fluxes $\Phi_{i = 1,2}$~\cite{Tovmasyan2018}. 
	
	For our purposes, it is useful to note that the average ($\vb{q} = \vb{0}$) current density operator for nonzero $\vb{A}$ is given by the relation
	\begin{equation}
		\label{eq:h_A_current}
		\hat{\vb{J}}(\vb{q} = \vb{0}) = -\frac{1}{\mathcal{A}} \grad_{\vb{A}} \mathcal{\hat H}_{\rm free}(\vb{A}) \,.
	\end{equation}
	This definition coincides with~\eqref{eq:Jq} for $\vb{A} = \vb{0}$.

	\section{Mean-field theory for superconducting systems from the Bogoliubov inequality}
	\label{sec:mean-field}
	
	The idea of mean-field theory is to approximate an interacting system by means of an effective noninteracting one. At zero temperature, this means finding among the wavefunctions that are represented as a single Slater determinant the one that minimizes the expectation value $\langle \hat{\mathcal{H}} \rangle$ of the many-body Hamiltonian~\eqref{eq:Ham_tot}. Indeed, mean-field theory is an application of the variational principle of quantum mechanics~\cite{Parravicini2013}. In the case of Hartree-Fock mean-field theory, the search is restricted to the space of wavefunctions with fixed particle number.  The key contribution of Bardeen, Cooper and Schrieffer~\cite{Bardeen1957,Bardeen1957a,Schrieffer1964,Tinkham2004} was to understand that mean-field theory can be successfully applied to superconducting systems, if one relaxes the condition of fixed particle number. As a consequence, anomalous expectation values of the form $\langle \hat{c}_i\hat{c}_j\rangle$ and $\langle \hat{c}_i^\dg\hat{c}_j^\dg\rangle$ can be nonzero and play the role of the order parameter of the superconductive state.

	In the following, we derive the self-consistency equations of BCS mean-field theory~\eqref{eq:BCS_self_cons_1}-\eqref{eq:BCS_self_cons_3} using the Bogoliubov inequality as a starting point. Beside the advantage of providing a self-contained and pedagogical exposition, there are two main reasons for using this approach: i) it emphasizes the fact that BCS theory is based on a rigorous variational principle, even at finite temperature. This is essential to justify the result for the  superfluid weight presented in Sec.~\ref{sec:superfluid_weight}. ii) It allows to recast the mean-field problem as the minimization of a suitable function of the one-particle density matrix, as detailed in Sec.~\ref{sec:mermin_functional}. Our derivation is similar to the one presented in Ref.~\onlinecite{Chan2022a}, but is more general since we consider a wider class of interaction terms and no unnecessary assumptions on the symmetries of the system are made (for instance time-reversal symmetry is not required).
	
	Consider a generic Hamiltonian which is the sum of two parts $\ham = \ham_0+\lambda\ham_1$, where $\lambda$ is a parameter. It is proven in Ref.~\onlinecite{Feynman1972} that the free energy $F(\lambda) = -\beta^{-1}\ln \Tr\qty\big[e^{-\beta(\ham_0+\lambda\ham_1)}]$, with $\beta = 1/(k_{\rm B}T)$ the inverse temperature, 
 	is a concave function of $\lambda$, that is $\dv*[2]{F(\lambda)}{\lambda} \leq 0$. Using the concavity property, it is easy to obtain the Bogoliubov inequality~\cite{Kuzemsky2015,Feynman1972,J.J.Binney1992}
 	\begin{equation}
 		\label{eq:Bogoliubov_ineq}
 		F \leq F_0 + \langle \ham -\ham_0\rangle\,,  
 	\end{equation}
	where $F_{(0)} = -\beta^{-1} \ln \mathcal{Z}_{(0)}= -\beta^{-1} \ln \Tr\big[e^{-\beta\ham_{(0)}}\big]$ and the expectation value on the right hand side is taken with respect to the Hamiltonian $\ham_0$, that is $\langle\, \cdot\,\rangle = \Tr\big[\cdot\,e^{-\beta \ham_0}\big]/\mathcal{Z}_0$. The left hand side of the inequality  is the free energy of the many-body system which we would like to approximate, the right hand side is the variational mean-field free energy which needs to be minimized to obtain the best possible approximation. This is similar to the zero temperature case, in which the optimization is carried out over all possible wavefunctions of noninteracting fermions. According to the variational principle, the best wavefunction is the one with the lowest energy since any energy expectation value taken on the many-body Hamiltonian is bound to be larger than the exact ground state energy. In fact, the variational principle can  be recovered from~\eqref{eq:Bogoliubov_ineq}: in the zero temperature limit the free energy becomes the ground state energy $F \to E_{\rm GS}$ and $F_0 \to E_{GS,0} = \expval*{\mathcal{\hat H}_0} = \expval{\mathcal{\hat H}_0}{\Psi_{{\rm GS},0}}$ where $\ket{\Psi_{{\rm GS},0}}$ is the ground state of $\mathcal{\hat H}_0$ (a nondegenerate ground state is assumed), thus one obtains $E_{\rm GS} \leq \expval*{\mathcal{\hat H}} = \expval{\mathcal{\hat H}}{\Psi_{{\rm GS},0}}$. Since $\mathcal{\hat H}_0$ can be an arbitrary Hamiltonian, the wavefunction $\ket{\Psi_{{\rm GS},0}}$ is also arbitrary and one recovers the variational principle. This shows that the Bogoliubov inequality is the right tool to generalize the variational principle to the finite temperature case.
	
	We apply the Bogoliubov inequality with $\mathcal{\hat H} = \mathcal{\hat H}_{\rm free} + \mathcal{\hat H}_{\rm int}-\mu\hat{N}$, where the noninteracting part $\mathcal{\hat H}_{\rm free}$ is given by~\eqref{eq:Ham_free} or~\eqref{eq:Ham_free_A}, the interaction term $\mathcal{\hat H}_{\rm int}$ by~\eqref{eq:Ham_int} and we have added a chemical potential term $-\mu \hat{N}$ ($\hat{N} =  \sum_{i} \hat{n}_{i}$ is the total particle number operator). The chemical potential is necessary to control the particle density in the case of Hamiltonians that do not conserve the particle number. This means that we are working in the grand canonical ensemble and the free energies $F,\,F_0$ appearing in~\eqref{eq:Bogoliubov_ineq} are in fact grand potentials $\Omega,\,\Omega_0$ and will be referred to as such from now on.
	
	For the variational Hamiltonian $\mathcal{\hat H}_0$ we take~\cite{Altland1997}
	\begin{equation}
		\label{eq:H_BdG}
		\mathcal{\hat H}_0 = \mathcal{\hat H}_{\rm free}-\mu\hat{N} + \sum_{i,j} \qty(\Gamma_{i,j} \hat{c}_{i}^\dg \hat{c}_{j} + \frac{1}{2}\Delta_{i,j} \hat{c}_{i}^\dg \hat{c}_{j}^\dg + \frac{1}{2}\Delta_{i,j}^*\hat{c}_{j}\hat{c}_{i})\,.
	\end{equation}
	The coefficients $\Gamma_{i,j}$ and $\Delta_{i,j}$ are variational parameters with the only constraints $\Gamma_{i,j}^* = \Gamma_{j,i}$, since $\mathcal{\hat H}_0$ must be Hermitian, and $\Delta_{i,j} = -\Delta_{j,i}$ because of the fermionic anticommutation relations $\lbrace \hat{c}_{i}^\dg, \hat{c}_{j}^\dg \rbrace = \lbrace \hat{c}_{i}, \hat{c}_{j} \rbrace = 0$.
	The last term on the right hand side of~\eqref{eq:H_BdG} is simply the most general quadratic Hamiltonian,  also including terms of the form $\hat{c}_{i}^\dg \hat{c}_{j}^\dg$ and $\hat{c}_{j}\hat{c}_{i}$ that break particle number conservation, which are crucial in superconducting systems. It is convenient to separate in $\mathcal{\hat H}_0$ the quadratic term $\mathcal{\hat H}_{\rm free}-\mu\hat{N}$ that appears also in the full many-body Hamiltonian. In this way the coefficients $\Gamma_{i,j}$ and $\Delta_{i,j}$ can be interpreted as effective fields that describe the averaged effect of interactions: $\Gamma_{i,j}$ is the Hartree-Fock potential and $\Delta_{i,j}$ is known as the pairing (or pair) potential~\cite{deGennes1966}.
	
	Since the variational Hamiltonian $\ham_0$ is quadratic, the expectation value $\langle \ham -\ham_0\rangle$ in~\eqref{eq:Bogoliubov_ineq} can be evaluated using Wick's theorem~\cite{Fetter2003}
	\begin{equation}
		\label{eq:inter_Bog}
		\begin{split}
			\expval*{\mathcal{\hat H}-\mathcal{\hat H}_0} &= 
			\frac{1}{2}\sum_{i,j} V_{i,j}\qty(\expval{\hat{n}_{i}}\expval{\hat{n}_{j}}-\expval*{\hat{c}_{i}^\dg\hat{c}_{j}}\expval*{\hat{c}_{j}^\dg\hat{c}_{i}} + \expval*{\hat{c}_{i}^\dg\hat{c}_{j}^\dg}\expval*{\hat{c}_{j}\hat{c}_{i}}) \\
			&\quad- \sum_{i,j} \qty(\Gamma_{i,j} \expval*{\hat{c}_{i}^\dg \hat{c}_{j}} + \frac{1}{2}\Delta_{i,j} \expval*{\hat{c}_{i}^\dg \hat{c}_{j}^\dg} + \frac{1}{2}\Delta_{i,j}^*  \expval*{\hat{c}_{j}\hat{c}_{i}})\,.
		\end{split}
	\end{equation}
	The minimization of the right hand size of~\eqref{eq:Bogoliubov_ineq} is most easily performed not with respect to the effective fields, but rather with respect to their conjugate variables, that is the expectation values of quadratic operators appearing in~\eqref{eq:inter_Bog}, which are obtained as derivatives of the grand potential $\Omega_0 = -\beta^{-1} \ln\Tr\big[e^{-\beta\mathcal{\hat H}_0}\big]$  (denoted by $F_0$ in Eq.~\eqref{eq:Bogoliubov_ineq}) associated to the Hamiltonian $\ham_0$
		\begin{gather}
			\pdv{\Omega_0}{\Gamma_{i,j}} = \expval*{\hat{c}_{i}^\dg\hat{c}_{j}}\,,
			\label{eq:omega_deriv_1}
			\\
			\pdv{\Omega_0}{\Delta_{i,j}} = \expval*{\hat{c}_{i}^\dg \hat{c}_{j}^\dg}\,,\qquad \pdv{\Omega_0}{\Delta^*_{i,j}} = \expval*{\hat{c}_{j}\hat{c}_{i}}\,,
			\label{eq:omega_deriv_2}
		\end{gather}
	as one can show using~\eqref{eq:Feynman_result} in Appendix~\ref{app:equivalence}. 
	Assuming that the mapping from the effective fields to the expectation values $\Gamma_{i,j},\,\Delta_{i,j},\,\Delta_{i,j}^* \to \expval*{\hat{c}_{i}^\dg\hat{c}_{j}},\,\expval*{\hat{c}_{i}^\dg \hat{c}_{j}^\dg},\,\expval*{\hat{c}_{j}\hat{c}_{i}}$, as given by~\eqref{eq:omega_deriv_1}-\eqref{eq:omega_deriv_2}, can be inverted, one can perform the Legendre transform of the grand potential $\Omega_0$, which is
	\begin{equation}
		\label{eq:Omega_0_bar}
		\overline{\Omega}_0  = \Omega_0 - \sum_{i,j} \qty(\Gamma_{i,j} \expval*{\hat{c}_{i}^\dg \hat{c}_{j}} + \frac{1}{2}\Delta_{i,j} \expval*{\hat{c}_{i}^\dg \hat{c}_{j}^\dg} + \frac{1}{2}\Delta_{i,j}^*  \expval*{\hat{c}_{j}\hat{c}_{i}})\,. 
	\end{equation}
	The Legendre transform $\overline{\Omega}_0$ is a function of the expectation values $\expval*{\hat{c}_{i}^\dg\hat{c}_{j}},\,\expval*{\hat{c}_{i}^\dg \hat{c}_{j}^\dg},\,\expval*{\hat{c}_{j}\hat{c}_{i}}$, which are the new independent variables, while the effective fields are now dependent variables obtained as the derivatives of $\overline{\Omega}_0$
	\begin{equation}
		\label{eq:omega_deriv_inv}
		\pdv{\overline{\Omega}_0}{\expval*{\hat{c}_{i}^\dg\hat{c}_{j}}} = -\Gamma_{i,j}\,,
		\qquad
		\pdv{\overline{\Omega}_0}{\expval*{\hat{c}_{i}^\dg\hat{c}_{j}^\dg}} = -\Delta_{i,j}
		\qquad
		\pdv{\overline{\Omega}_0}{\expval*{\hat{c}_{j}\hat{c}_{i}}} = -\Delta^*_{i,j}\,.
	\end{equation}
	We have now reduced the problem to minimizing the function $\Omega_{\rm m.f.} \qty\big(\expval*{\hat{c}_{i}^\dg\hat{c}_{j}},\expval*{\hat{c}_{i}^\dg \hat{c}_{j}^\dg},\expval*{\hat{c}_{j}\hat{c}_{i}})$ on the right hand side of~\eqref{eq:Bogoliubov_ineq} (the mean-field grand potential), given by 
	\begin{equation}
		\label{eq:f_def}
		\begin{split}
		\Omega_{\rm m.f.}\big(\expval*{\hat{c}_{i}^\dg\hat{c}_{j}},\expval*{\hat{c}_{i}^\dg \hat{c}_{j}^\dg},\expval*{\hat{c}_{j}\hat{c}_{i}}) &= \overline{\Omega}_0\big(\expval*{\hat{c}_{i}^\dg\hat{c}_{j}},\expval*{\hat{c}_{i}^\dg \hat{c}_{j}^\dg},\expval*{\hat{c}_{j}\hat{c}_{i}}) 
		\\
		&+ 
		\frac{1}{2}\sum_{i,j} V_{i,j}\qty(\expval{\hat{n}_{i}}\expval{\hat{n}_{j}}-\expval*{\hat{c}_{i}^\dg\hat{c}_{j}}\expval*{\hat{c}_{j}^\dg\hat{c}_{i}} + \expval*{\hat{c}_{i}^\dg\hat{c}_{j}^\dg}\expval*{\hat{c}_{j}\hat{c}_{i}})\,.
		\end{split}
	\end{equation}
	By setting  the partial derivatives of $\Omega_{\rm m.f.}$ to zero and using~\eqref{eq:omega_deriv_inv} one obtains
	\begin{gather}
		\label{eq:BCS_self_cons_1}
		\Gamma_{i,i} =  \sum_{j} V_{i,j}\expval{\hat{n}_{j}}\,,\\
		\label{eq:BCS_self_cons_2}
		\Gamma_{i,j} = - V_{i,j}\expval*{\hat{c}_{j}^\dg\hat{c}_{i}}\,, \qq{for} i \neq j\,,\\
		\label{eq:BCS_self_cons_3}
		\Delta_{i,j} = V_{i,j}\expval*{\hat{c}_{j}\hat{c}_{i}}\,,\quad \Delta^*_{i,j} = V_{i,j}\expval*{\hat{c}_{i}^\dg\hat{c}_{j}^\dg}\,,\qq{for} i \neq j\,.
	\end{gather}
	These are known as the self-consistency equations of BCS mean-field theory.
	
	It is convenient at this point to define the mean-field Hamiltonian $\mathcal{\hat H}_{\rm m.f.}$ as
	\begin{equation}
		\label{eq:mean-field_ham}
		\mathcal{\hat H}_{\rm m.f.} = \mathcal{\hat H}_0 + \expval*{\mathcal{\hat H}-\mathcal{\hat H}_0}\hat{1}\,,
	\end{equation}
	where, as a reminder, the expectation value is taken with respect to the density matrix $\hat{\rho}_0 = e^{-\beta \mathcal{\hat H}_0}/\mathcal{Z}_0$. On the other hand, $\mathcal{\hat H}_{\rm m.f.}$ and $\mathcal{\hat H}_0$ differ by a constant, therefore all averages taken with respect to the statistical ensemble defined by $\mathcal{\hat H}_{\rm m.f.}$ are the same as the averages  taken with respect to $\mathcal{\hat H}_0$. In other words, the two Hamiltonians give the same density matrix $\hat{\rho}_0 =  e^{-\beta \mathcal{\hat H}_{\rm m.f.}}/\Tr\big[e^{-\beta \mathcal{\hat H}_{\rm m.f.}}\big]$. For this reason, we do not introduce any new notation to indicate the averages taken with respect to $\mathcal{\hat H}_{\rm m.f.}$. The mean-field Hamiltonian has the property that
	\begin{equation}
		\label{eq:expval_H_Hmf}
		\langle \mathcal{\hat H} \rangle = \langle \mathcal{\hat H}_{\rm m.f.} \rangle\,.
	\end{equation}
	Using the mean-field Hamiltonian~\eqref{eq:mean-field_ham}, the Bogoliubov inequality~\eqref{eq:Bogoliubov_ineq} can be rewritten as
	\begin{equation}
		\label{eq:Bogoliubov_second}
		\Omega \leq \Omega_{\rm m.f.} = -\beta^{-1}\ln \Tr\big[e^{-\beta \mathcal{\hat H}_{\rm m.f.}}\big]\,,	
	\end{equation}
	with $\Omega_{\rm m.f.}$ the mean-field grand potential introduced in~\eqref{eq:f_def}.
	These last two equations provide a better understanding of the finite temperature variational principle: the variational mean-field Hamiltonian $\mathcal{\hat H}_{\rm m.f.}$ is optimized so as to minimize the grand potential $\Omega_{\rm m.f.}$ with the only constraint being that the expectation value of the many-body Hamiltonian $\mathcal{\hat H}$ computed with respect to the statistical ensemble defined by $\mathcal{\hat H}_{\rm m.f.}$ is the same as the expectation value $\langle \mathcal{\hat H}_{\rm m.f.} \rangle$, which is computed with the same ensemble.

	\section{Mermin-Gibbs functional for superconducting systems and linear scaling methods}
	\label{sec:mermin_functional}
	
	In the previous section, mean-field theory for superconductive systems has been formulated as the minimization of the mean-field grand potential  $\Omega_{\rm m.f.}$~\eqref{eq:f_def}, which, after a Legendre transform, becomes a function of the expectation values $\expval*{\hat{c}_{i}^\dg\hat{c}_{j}},\,\expval*{\hat{c}_{i}^\dg \hat{c}_{j}^\dg},\,\expval*{\hat{c}_{j}\hat{c}_{i}}$. These expectation values replace the effective fields $\Gamma_{i,j}$ and $\Delta_{i,j}$ as the variational parameters. This change of variables has been useful to derive the self-consistency equations in a straightforward manner. The goal of this section is to explicitly write down the mean-field grand potential as a function of the expectation values $\expval*{\hat{c}_{i}^\dg\hat{c}_{j}},\,\expval*{\hat{c}_{i}^\dg \hat{c}_{j}^\dg},\,\expval*{\hat{c}_{j}\hat{c}_{i}}$ and identify the domain of this function. This is the first main result of this work and it is important in order to have a mathematically well-defined minimization problem, which is the starting point for the application of linear scaling methods. We start by reviewing the Nambu space formalism.
	
	The idea of Nambu was to write the variational Hamiltonian $\ham_0$~\eqref{eq:H_BdG} in the \qql row multiplies matrix multiplies
	column\qqr format~\cite{Schrieffer1964,Altland1997}
	\begin{equation}
		\label{eq:BdG}
		\begin{split}
			\mathcal{\hat H}_0 &= \frac{1}{2}\pmqty{\hat{\vb{c}}^\dg & \hat{\vb{c}}^T}
			\pmqty{K-\mu+\Gamma & \Delta \\ -\Delta^* & -\qty(K-\mu +\Gamma)^T } \pmqty{\hat{\vb{c}} \\(\hat{\vb{c}}^\dg)^T}+ \frac{1}{2}\Tr[K-\mu 1 +\Gamma]\,,
		\end{split}
	\end{equation}
	where $K$, $\Delta$ and $\Gamma$ are square matrices with matrix elements $K_{i,j}$, $\Delta_{i,j}$ and $\Gamma_{i,j}$, respectively, $\mu$ is the chemical potential and $\hat{\vb{c}}$ ($\hat{\vb{c}}^\dg$) is the column (row) vector of annihilation (creation) operators $\hat{c}_{i}$ ($\hat{c}_{i}^\dg$). We indicate with $A^T$ the transpose of the matrix $A$ and with $\vb{v}^T$ the row vector which is the transpose of the column vector $\vb{v}$.
	
	The single-particle Hamiltonian appearing in~\eqref{eq:BdG}
	\begin{equation}
		\label{eq:H_0_BdG}
		H_0 = \pmqty{K-\mu +\Gamma & \Delta \\ -\Delta^* & -\qty(K-\mu +\Gamma)^T } = \pmqty{h & \Delta \\ -\Delta^* & -h^{T} }\,.
	\end{equation}
	is known as the Bogoliubov-de Gennes (BdG) Hamiltonian~\cite{deGennes1966} and in general it has no constraints other than $h = K-\mu +\Gamma = h^\dg$ and $\Delta = -\Delta^T$. The BdG Hamiltonian acts on an enlarged Hilbert space called the particle-hole space, or alternatively the Nambu space. This space is the tensor product of the original single-particle Hilbert space with the so-called particle-hole degree of freedom. 
	The block structure of the BdG Hamiltonian can be conveniently summarized by the condition
	\begin{equation}
		\label{eq:particle_hole_1}
		H_0 = -\Sigma_x H_0^T \Sigma_x\,,\qq{with} \Sigma_x = \sigma_x \otimes 1 = \pmqty{0 & 1 \\ 1 & 0}\otimes 1\,,
	\end{equation}
	together with $H_0 = H_0^{\dg}$. Here $\Sigma_x$ is a unitary matrix that  acts as the $\sigma_x$ Pauli matrix on the particle-hole degree of freedom. 
	
	The BdG Hamiltonian can be diagonalized by a unitary transformation $G$ that respects the block structure encoded by~\eqref{eq:particle_hole_1}~\cite{Altland1997}. Specifically, we have $H_0 = G^\dg DG$ with
	\begin{gather}
		D  = \pmqty{\mathrm{diag}(E_i) & 0 \\ 0  & -\mathrm{diag}(E_i)}\,,\\
		G^{\dagger} = G^{-1} \qq{and} G = \Sigma_x (G^{-1})^T\Sigma_x\,.
	\end{gather}
	We use the unitary transformation $G$ to introduce new field operators $\hat{d}_i$ as
	\begin{equation}
		\label{eq:G_matrix_def}
		\pmqty{\hat{\vb{d}} \\ (\hat{\vb{d}}^\dagger)^T} = G\pmqty{\hat{\vb{c}} \\(\hat{\vb{c}}^\dagger)^T}\,,
	\end{equation}
	and from~\eqref{eq:BdG} we obtain 
	\begin{equation}
		\label{eq:BdG_Ham_diagonal}
		\mathcal{\hat H}_0 = \sum_{i}E_i\hat{d}_i^\dagger\hat{d}_i-\frac{1}{2}\sum_{i}E_i + \frac{1}{2}\Tr[K-\mu +\Gamma]\,.
	\end{equation}
	From this last result, we see that the eigenvalues $E_i$ of $H_0$ can be interpreted as excitation energies of fermionic quasiparticles on top of the mean-field ground state. Using~\eqref{eq:BdG_Ham_diagonal}, the grand potential $\Omega_0$ becomes
	\begin{equation}
		\label{eq:Omega_Ei}
		\Omega_0 = -\beta^{-1}\sum_i\ln(1+e^{-\beta E_i}) -\frac{1}{2}\sum_{i}E_i + \frac{1}{2}\Tr[K-\mu +\Gamma]\,.
	\end{equation}
	On the other hand, it is not difficult to prove the identity
	\begin{equation}
		-\frac{1}{2\beta}\Tr\ln(1+e^{-\beta H_0}) =  -\beta^{-1} \sum_i \ln\qty\big(1+e^{-\beta E_i})-\frac{1}{2}\sum_iE_i\,.
	\end{equation}
	Combining this last result with~\eqref{eq:Omega_Ei} leads to
	\begin{equation}
		\label{eq:Omega_0_BdG}
		\Omega_0 = -\frac{1}{2\beta}\Tr\ln(1+e^{-\beta H_0}) + \frac{1}{2}\Tr[K-\mu +\Gamma]\,.
	\end{equation}
	In this way the grand potential $\Omega_0$ has been conveniently expressed in terms of the single-particle BdG Hamiltonian $H_0$. As a consequence, \eqref{eq:omega_deriv_1},~\eqref{eq:omega_deriv_2} and~\eqref{eq:Omega_0_BdG} together give
	\begin{gather}
		\label{eq:exp_val_1_Omega0}
		\begin{split}
			\expval*{\hat{c}_{i}^\dg\hat{c}_{j}}  = \frac{1}{2}\Tr[\frac{1}{e^{\beta H_0}+1}\pdv{H_0}{\Gamma_{i,j}}]+ \frac{\delta_{i,j}}{2}\,,
		\end{split} \\
		\label{eq:exp_val_2_Omega0}
		\expval*{\hat{c}_{i}^\dg \hat{c}_{j}^\dg} = 
		 \frac{1}{2}\Tr[\frac{1}{e^{\beta H_0}+1}\pdv{H_0}{\Delta_{i,j}}]\,,\quad \expval*{\hat{c}_{j}\hat{c}_{i}}  = \frac{1}{2}\Tr[\frac{1}{e^{\beta H_0}+1}\pdv{H_0}{\Delta^*_{i,j}}]\,.
	\end{gather}
	Thus, it is clearly possible to express these expectation values in terms of the matrix elements of the single-particle operator $P = (e^{\beta H_0}+1)^{-1}$. The eigenvalues $p_i$ of $P$ are the occupation numbers of the eigenstates of the BdG Hamiltonian, which in the fermionic case satisfy the inequalities $0 \leq p_i \leq 1$. 
	One can equivalently express this condition as the matrix inequalities $0 \leq P \leq 1$, meaning that $0 \leq \ev{P}{\psi} \leq 1$ for any normalized single-particle state $\ket{\psi}$. 
	Moreover, $P$ is constrained by the particle-hole symmetry of $H_0$~\eqref{eq:particle_hole_1}, which leads to
	\begin{equation}
		\label{eq:P_symmetry}
		P = 1-\Sigma_x P^T \Sigma_x.
	\end{equation}
	Together with $P = P^{\dagger}$ this implies
	\begin{equation}
		\label{eq:P_block}
		P = \mqty(P_{11} & P_{12} \\ P_{21} & P_{22}) = \mqty(P_{11} & P_{12} \\ -P^*_{12} & 1-P_{11}^T)	\,,\qq{with} P_{11} =P_{11}^\dagger \qq{and} P_{12} = -P_{12}^T\,.
	\end{equation}
	From~\eqref{eq:exp_val_1_Omega0},~\eqref{eq:exp_val_2_Omega0} and~\eqref{eq:P_block} one obtains the useful relations
	\begin{equation}
		\label{eq:expval_P_relations}
		\expval*{\hat{c}_{i}^\dagger\hat{c}_{j}} = [P_{11}]_{j,i}\,,\qquad \expval*{\hat{c}_{i}^\dagger \hat{c}_{j}^\dagger} = [P_{21}]_{j,i}\,, \qquad  \expval*{\hat{c}_{i}\hat{c}_{j}} = [P_{12}]_{j,i}\,. 
	\end{equation}
	Moreover, the relation 
	\begin{equation}
		\label{eq:exp_val_to_eff_map}
		\mqty(\Gamma & \Delta \\
		-\Delta^* & -\Gamma^T) = \beta^{-1}\ln(P^{-1}-1) -\mqty(K -\mu & 0 \\ 0 & -(K^T -\mu))\,,
	\end{equation} 
	gives explicitly the mapping $ \expval*{\hat{c}_{i}^\dagger\hat{c}_{j}},\expval*{\hat{c}_{i}^\dagger \hat{c}_{j}^\dagger},\,\expval*{\hat{c}_{j} \hat{c}_{i}} \to \Gamma_{i,j},\,\Delta_{i,j},\,\Delta_{i,j}^*$, which has been used to perform the Legendre transform of the grand potential $\Omega_0$ in~\eqref{eq:Omega_0_bar}. Using this last result one can easily show that the mapping is one-to-one, therefore the Legendre transform is well-defined.
	
	In order to obtain the Legendre transform $\overline{\Omega}_0(P)$ as an explicit function of $P$, we write the second term on the right hand side of~\eqref{eq:Omega_0_bar}  as
	\begin{equation}
		\label{eq:Trace_eff_fields_P}
		\sum_{i,j} \qty(\Gamma_{i,j} \expval*{\hat{c}_{i}^\dagger \hat{c}_{j}} + \frac{1}{2}\Delta_{i,j} \expval*{\hat{c}_{i}^\dagger \hat{c}_{j}^\dagger} + \frac{1}{2}\Delta_{i,j}^*  \expval*{\hat{c}_{j}\hat{c}_{i}}) = \frac{1}{2}\Tr[\mqty(\Gamma & \Delta \\
		-\Delta^* & -\Gamma^T)P] + \frac{1}{2}\Tr[\Gamma]\,.
	\end{equation}
	Therefore, by combining~\eqref{eq:Omega_0_bar},~\eqref{eq:Omega_0_BdG},~\eqref{eq:exp_val_to_eff_map} and~\eqref{eq:Trace_eff_fields_P} and using $-\ln(1+e^{-\beta H_0}) = \ln (1-P)$, we obtain
	\begin{equation}
		\label{eq:Omega_0_bar_P_final}
		\begin{split}
			\overline{\Omega}_0(P) &= \frac{1}{2\beta}\Tr[P\ln P + (1-P)\ln (1-P)] +\Tr[(K-\mu)P_{11}] \\
			&= -TS(P)+\Tr[(K-\mu)P_{11}]\,.
		\end{split}
	\end{equation}
	Here, we have identified the (von Neumann) entropy $S(P) = -\frac{k_{\rm B}}{2}\Tr[P\ln P + (1-P)\ln (1-P)]$ of a gas of noninteracting fermionic (quasi-)particles (note the factor $1/2$, which is absent in the usual definition of the von Neumann entropy). Thus minimizing $\overline{\Omega}_0$ is the same as maximizing the entropy with the constraint that the expectation value of the energy (the second term in~\eqref{eq:Omega_0_bar_P_final}) is fixed. Indeed, the inverse temperature $\beta = (k_{\rm B}T)^{-1}$ is the Lagrange multiplier corresponding to the energy constraint.
	In the presence of interactions, the function to be minimized is (see~\eqref{eq:f_def})
	\begin{equation}
		\label{eq:Omega_LWBK}
		\begin{split}
			\Omega_{\rm m.f.}(P) &= \overline{\Omega}_0(P) +  \frac{1}{2}\sum_{i,j} V_{i,j}\qty(\expval{\hat{n}_{i}}\expval{\hat{n}_{j}}-\expval*{\hat{c}_{i}^\dagger\hat{c}_{j}}\expval*{\hat{c}_{j}^\dagger\hat{c}_{i}} + \expval*{\hat{c}_{i}^\dagger\hat{c}_{j}^\dagger}\expval*{\hat{c}_{j}\hat{c}_{i}}) \\
			&=\overline{\Omega}_0(P) + V(P)\,,
		\end{split}
	\end{equation}
	where $V(P) = \langle \ham_{\rm int}\rangle$ is the quadratic form defined as the second term in the first line. The minimization of $\Omega_{\rm m.f.}(P)$ has to be carried out within the domain of single-particle operators $P$ which satisfy the conditions $0 \leq P \leq 1$ and the symmetry constraint~\eqref{eq:P_symmetry}. At zero temperature ($\beta = +\infty$) the entropy is zero, namely the eigenvalues of $P$ can be only zero or one, therefore one has the idempotency constraint $P^2 = P$ on the one-particle density matrix. 
	
	From~\eqref{eq:Omega_0_bar_P_final} and~\eqref{eq:Omega_LWBK} it is clear that gauge invariance is respected by the mean-field approximation. Indeed, under a gauge transformation~\eqref{eq:gauge_transf} a single-particle operator $O$, corresponding to a quadratic many-body operator $\hat{O} = \sum_{i,j}\hat{c}_{i}^\dagger O_{i,j}\hat{c}_{j}$, transforms as
	\begin{equation}
		O' = U_{\rm g}OU_{\rm g}^{\dagger}\,,\qq{with} [U_{\rm  g}]_{i,j} = \delta_{i,j}e^{-i\theta_{i}}\,.
	\end{equation}
	It is then evident that both $\overline{\Omega}_0$ and $\Omega_{\rm m.f.}$ are invariant under the transformations $K \to K' = U_{\rm  g}KU_{\rm g}^{\dagger}$ and 
	\begin{equation}
		P \to P' = \mqty(U_{\rm  g} & 0 \\ 0 & U_{\rm  g}^{\dagger})P\mqty(U_{\rm  g}^\dagger & 0 \\ 0 & U_{\rm  g})\,.
	\end{equation}
	
	The explicit form of the mean-field grand potential $\Omega_{\rm m.f.}(P)$ as a function of the single-particle density matrix $P$ as given by~\eqref{eq:Omega_0_bar_P_final} and~\eqref{eq:Omega_LWBK} is one of the main results of this work and serves as the starting point for the application of linear scaling methods for superconducting systems. Here, we have carried out the derivation in the case of a BdG Hamiltonian of the form in~\eqref{eq:H_0_BdG}, which is the most general structure that a BdG Hamiltonian can have. It is straightforward to specialize~\eqref{eq:Omega_0_bar_P_final}-\eqref{eq:Omega_LWBK} to systems with additional symmetries, such as time-reversal symmetry, which belong to different classes within the topological classification of noninteracting fermionic Hamiltonians~\cite{Schnyder2008,Chiu2016}.  
	
	The result in~\eqref{eq:Omega_0_bar_P_final}-\eqref{eq:Omega_LWBK} is aesthetically appealing since it takes the same form as the Mermin-Gibbs functional for a system of noninteracting fermions~\cite{Mermin1965,Corkill1996,MichaelLindsey2019} with the only difference that the single-particle density matrix $P$ includes the anomalous expectation values $\expval*{\hat{c}_{i}^\dg \hat{c}_{j}^\dg},\,\expval*{\hat{c}_{j}\hat{c}_{i}}$ in the off-diagonal blocks~\eqref{eq:expval_P_relations}. Linear scaling method are generally implemented at zero temperature, in which case the density matrix satisfies the idempotency condition $P^2 = P$. Ensuring that the minimization of the total  energy is performed within the space of idempotent density matrices is the main technical difficulty and a number of ingenious techniques have been developed to address this problem~\cite{Goedecker1999,Bowler2012}. In the case of superconducting systems one would also need to enforce the additional constraint given by particle-hole symmetry~\eqref{eq:P_symmetry}. However, this is a much simpler constraint to deal with compared to idempotency since it simply reflects a redundancy in the matrix elements of the density matrix. For this reason, we expect that all the linear scaling methods developed so far for normal (non-superconducting) system can be readily implemented and used for superconducting systems as well. Linear scaling methods have been employed also at nonzero temperature in a few cases~\cite{Corkill1996} and again we do not see any fundamental difficulty in doing the same for superconducting systems. Being able to work at finite temperature is obviously important in the context of superconductivity.
	
	The key fact behind the success of linear scaling methods is that the one-particle density matrix is ranged~\eqref{eq:density_matrix_decay}. There are no available results regarding the decay of the density matrix with distance in the case of superconducting systems. However, given that the BdG Hamiltonian $H_0$ is generally gapped, it is reasonable to expect an exponential decay as in insulators. The important problem of the localization properties of the density matrix in superconducting systems will be the subject of future studies. We note in passing that superconductivity is the manifestation of off-diagonal long range-order in the two-particle density matrix. In the case of a conventional $s$-wave superconductor this means $\expval*{\hat{c}_{i\uparrow}^\dg\hat{c}_{i\downarrow}^\dg\hat{c}_{j\downarrow}\hat{c}_{j\uparrow}} \to \qty\big|\expval*{\hat{c}_{i\uparrow}^\dg\hat{c}_{i\downarrow}^\dg}|^2\neq 0$ for $|\vb{r}_i-\vb{r}_j| \to \infty$ (note that we have introduced the spin here). Thus the anomalous expectation values keep track of this off-diagonal long-range order, while there is no off-diagonal long-range order in the one-particle density matrix. Any form of  off-diagonal long-range order in $P$ would hinder the application of linear scaling methods, which are based on the approximation of setting  the matrix elements $P_{i,j}$ to zero,  if $d(i,j) > R$ for a given range $R$.
	
	\section{Superfluid weight and generalized random phase approximation}
	\label{sec:superfluid_weight}


In this Section, we present one of the main results of this work, namely we show that taking the full derivatives of the mean-field grand potential with respect to the constant vector potential $\vb{A}$, which implements twisted boundary conditions (see Section~\ref{sec:basics}), leads to the superfluid weight in the generalized random phase approximation. The derivation is based on the results of the previous Section~\ref{sec:mermin_functional}, in particular the explicit expression for the mean-field grand potential as a function of the one-particle density matrix~\eqref{eq:Omega_0_bar_P_final}-\eqref{eq:Omega_LWBK}. We first explain how the superfluid weight can be calculated in different but equivalent ways.
	
The superfluid weight is defined in terms of a response function $\chi_{lm}(\vb{q},\vb{q}',\omega)$, that of the current
induced in the system to linear order in the vector potential~\cite{Schrieffer1964,Giuliani2005}
\begin{equation}
	\label{eq:chi}
J_{l}(\vb{q},\omega) = \sum_{m,\vb{q}'}\chi_{lm}(\vb{q},\vb{q}',\omega)A_m(\vb{q}',\omega)\,.	
\end{equation}
Here $\vb{J}(\vb{r},t) = \sum_{\vb{q}}\int\frac{\dd{\omega}}{2\pi}
\vb{J}(\vb{q},\omega)e^{i(\vb{q}\cdot\vb{r}-\omega t)}$ is the current density  with $\vb{J}(\vb{q},\omega)$ its Fourier transform, and similarly $\vb{A}(\vb{q},\omega)$ is the Fourier transform of the vector potential $\vb{A}(\vb{r},t)$.
We do not assume translational invariance here, therefore the response function depends on two wavevectors ($\vb{q}$, $\vb{q}'$). The response function is the sum of two parts $\chi_{lm} = \chi^{\rm d}_{lm} + \chi^{\rm p}_{lm}$, the diamagnetic response $\chi^{\rm d}_{lm}$ and the paramagnetic response $\chi^{\rm p}_{lm}$, given by (in the long-wavelength limit)
	\begin{gather}
		\chi_{lm}^{\rm d}(\vb{q},\vb{q}') =\frac{1}{\mathcal{A}}\sum_{i,j}[\vb{r}_{i,j}]_l[\vb{r}_{i,j}]_m\langle \hat{c}_{i}^\dagger K_{i,j} \hat{c}_{j}\rangle e^{-i(\vb{q}-\vb{q}')\cdot\frac{\vb{r}_{i}+\vb{r}_{j}}{2}}\,, \label{eq:chi_d}\\
		\chi_{lm}^{\rm p}(\vb{q},\vb{q}',\omega) = i\mathcal{A}\int^{+\infty}_{0} \dd{t} e^{i\omega t}
		\big\langle \qty\big[\hat{J}_{l}(\vb{q},t),\hat{J}_{m}(-\vb{q}',0)]\big\rangle\,. \label{eq:chi_p}
	\end{gather}
	In the above equations $\hat{J}_{l}(\vb{q},t) = e^{i\ham t}\hat{J}_{l}(\vb{q})e^{-i\ham t}$ is the $l = x,y$ component of the current operator~\eqref{eq:Jq} in the Heisenberg picture, while we denote with $[\vb{r}_{i,j}]_l$ the $l$ component of the displacement vector $\vb{r}_{i,j}$, which is defined in the caption of Figure~\ref{fig:lattice}. 
	In~\eqref{eq:chi_d} and~\eqref{eq:chi_p} the expectation values are taken with respect to the many-body Hamiltonian, that is $\langle\, \cdot\,\rangle = \Tr\big[\cdot\,e^{-\beta \ham}\big]/\mathcal{Z}$, contrary to the convention of Sections~\ref{sec:mean-field}-\ref{sec:mermin_functional}. 
	The superfluid weight is defined as the following limit of the response function
	\begin{equation}
		\label{eq:superfluid_weight}
		D_{{\rm s},lm} = -\lim_{\vb{q}'_\perp\to \vb{0}} \chi_{lm}(\vb{q}=\vb{0},q_\parallel'=0,\vb{q}'_\perp,\omega = 0)\,,
	\end{equation}
	where the wavevector $\vb{q}' = q'_\parallel \hat{m} + \vb{q}'_\perp$ is decomposed into the collinear ($q_{\parallel}'$) and perpendicular ($\vb{q}_\perp'$) components with respect to the $m = x,y$ axis ($m$ is the second index appearing in both $D_{{\rm s},lm}$ and $\chi_{lm}$ and $\hat{m}$ is the unit vector along the same direction). As a consequence of gauge invariance, the alternative limit of the response function is zero~\cite{Schrieffer1964,Scalapino1993}
	\begin{equation}
		\label{eq:gauge_constrain}
			 \lim_{q_\parallel'\to 0} \chi_{lm}(\vb{q}=\vb{0},q_\parallel',\vb{q}'_\perp=\vb{0},\omega = 0) = 0\,.
	\end{equation}
	The fact that these two limits do not commute is a consequence of the presence of long-range correlations in the current-current response function~\cite{Scalapino1993}.
	
	An equivalent definition of the superfluid weight is as the second derivative of the grand potential with respect to the fluxes $\Phi_i$ parametrizing twisted boundary conditions~\cite{Fisher1973}. As discussed in  Sec.~\ref{sec:basics}, twisted boundary conditions are equivalent to a constant vector potential $\vb{A}$ that cannot be gauged away for periodic boundary conditions, thus we have
	\begin{equation}
	\label{eq:d2F_dA2}
	 D_{{\rm s},lm} = \frac{1}{\mathcal{A}}\pdv{\Omega(\vb{A})}{A_l}{A_m}\bigg|_{\vb{A}=\vb{0}}\,,
	\end{equation}
	where $\Omega(\vb{A}) = -\beta^{-1}\ln\Tr\qty\big[e^{-\beta\ham(\vb{A})}]$, $\ham(\vb{A})  = \ham_{\rm free}(\vb{A}) +\ham_{\rm int}$  and $\ham_{\rm free}(\vb{A})$ is given by~\eqref{eq:Ham_free_A}. It is understood in the above definition that $\Omega(\vb{A})$ is the grand potential of a finite system with area $\mathcal{A}$ and that the thermodynamic limit $(\mathcal{A}\to \infty)
	$ is performed after the derivative is taken. The equivalence of the two definitions~\eqref{eq:superfluid_weight} and~\eqref{eq:d2F_dA2} is discussed for instance in Refs.~\onlinecite{Scalapino1993} and~\onlinecite{Taylor2006}. There is ambiguity in the literature as whether one should use the free energy (fixed particle number) or the grand potential (fixed chemical potential) in~\eqref{eq:d2F_dA2}. The two definitions usually lead to the same result, but this is not guaranteed in general. For simplicity, we choose to use the grand potential here and refer to Ref.~\onlinecite{Huhtinen2022} for an in depth discussion regarding this issue. 
	
	Evaluating the exact response function~\eqref{eq:chi}-\eqref{eq:chi_p} is a very difficult task, therefore one has to resort to  approximations such as the mean-field approximation presented in Section~\ref{sec:mean-field}. The two equivalent definitions~\eqref{eq:superfluid_weight} and~\eqref{eq:d2F_dA2} suggest two possible strategies for evaluating the superfluid weight within the mean-field approximation. The first one is to replace the many-body Hamiltonian $\ham$ with the mean-field Hamiltonian $\ham_{\rm m.f.}$ when evaluating the expectation values in~\eqref{eq:chi_d}-\eqref{eq:chi_p}. This approach (or equivalent ones) has been used in Ref.~\onlinecite{Scalapino1993} (see Sec.~III in this reference) and Refs.~\onlinecite{Liang2017a},~\onlinecite{Peotta2015}, but has the disadvantage that the response function does not respect gauge invariance since the gauge constraint~\eqref{eq:gauge_constrain} is not satisfied~\cite{Scalapino1993}. Moreover, the procedure of replacing the many-body Hamiltonian with the mean-field Hamiltonian in the response function is a rather uncontrolled approximation since the latter Hamiltonian is simply a collection of variational parameters, as explained in Section~\ref{sec:mean-field}, and there is no a priori reason to use it to compute response functions. Given that mean-field theory can be formulated as a variational approximation for the grand potential, a more justified approach consists in replacing the exact grand potential $\Omega(\vb{A})$ in~\eqref{eq:d2F_dA2} with its mean-field approximation $\Omega_{\rm m.f.}(\vb{A})$. The second main result of this work is to show that this approach leads to a well-known gauge-invariant approximation,
	the generalized random phase approximation.

	The effective fields $\Gamma$ and $\Delta$ are chosen so as to minimize $\Omega_{\rm m.f.}(\vb{A},\Gamma,\Delta)$ for a given value of $\vb{A}$, therefore they become themselves functions of the vector potential. This means that the mean-field grand potential depends on $\vb{A}$ either directly through the noninteracting Hamiltonian $\ham_{\rm free}(\vb{A})$ in~\eqref{eq:H_BdG} or indirectly through the effective fields. Using~\eqref{eq:d2F_dA2}, we have that an approximation for the superfluid weight is given by
	\begin{equation}
		\label{eq:D_mf}
		D_{{\rm s},lm} \approx 
		\frac{1}{\mathcal{A}}
		\frac{d^2\Omega_{\rm m.f.}\big(\vb{A},\Gamma(\vb{A}),\Delta(\vb{A})\big)}{dA_l dA_m}
		\bigg|_{\vb{A}=\vb{0}}\,.
	\end{equation}
	In this equation and in the following, we denote with $d/dA_l$ the full derivative with respect to $\vb{A}$, including both the direct and the indirect dependence, while the partial derivative $\pdv*{\Omega_{\rm m.f.}\big(\vb{A},\Gamma(\vb{A}),\Delta(\vb{A})\big)}{A_l}$ denotes the derivative with respect to the first argument only (direct dependence). We note that replacing the full derivative $\dv*{}{A_l}$ with the partial derivative $\pdv*{}{A_l}$ in~\eqref{eq:D_mf}, as done for instance in Ref.~\onlinecite{Peotta2015}, is the same as using the response functions~\eqref{eq:chi_d}-\eqref{eq:chi_p} computed with the mean-field Hamiltonian (see Appendix~\ref{app:equivalence}). In two recent works~\cite{Chan2022a,Huhtinen2022} it has been pointed out that it is important to take into account the dependence of the effective fields on the vector potential $\vb{A}$ when using~\eqref{eq:D_mf}, otherwise one may obtain unphysical results in the case of multiorbital lattices, in particular when flat bands are present. We mention in passing that  in Ref.~\onlinecite{Huhtinen2022} an alternative approach based on a modified form of linear response theory and equivalent to~\eqref{eq:D_mf} has been proposed.

	In Ref.~\onlinecite{Huhtinen2022} it has been shown that it is possible evaluate~\eqref{eq:D_mf} by solving the mean-field problem for $\vb{A} = \vb{0}$ only. The basic idea is the following: the variational parameters $\Gamma(\vb{A})$ and $\Delta(\vb{\vb{A}})$ are chosen so as to minimize the mean-field grand potential, as a consequence
	\begin{equation}
		\label{eq:minimizer}
		\pdv{\Omega_{\rm m.f.}\qty\big(\vb{A},\Gamma,\Delta)}{\Gamma}\bigg|_{\substack{\Gamma = \Gamma(\vb{A})\\ \Delta = \Delta(\vb{A})}} = 
		\pdv{\Omega_{\rm m.f.}\qty\big(\vb{A},\Gamma,\Delta)}{\Delta}\bigg|_{\substack{\Gamma = \Gamma(\vb{A})\\ \Delta = \Delta(\vb{A})}} = 0\,.
	\end{equation} 
	Taking the total derivative with respect to $\vb{A}$ of these equations leads to implicit equations for the derivatives $\pdv*{\Gamma(\vb{A})}{\vb{A}}$ and $\pdv*{\Delta(\vb{A})}{\vb{A}}$, which require only the knowledge of correlation functions at $\vb{A} =\vb{0}$. Using this approach and the results of Ref.~\onlinecite{Huhtinen2022} as a starting point it is possible to arrive to our final result~\eqref{eq:main_Ds_GRPA} after some cumbersome algebra.
	We do not present the details of the derivation here since it is simpler to use the same idea in the case where the mean-field grand potential is a function of the one-particle density matrix $P(\vb{A})$, namely $\Omega_{\rm m.f.}(\vb{A},P(\vb{A}))$ in~\eqref{eq:D_mf}, as shown in the following.
	
	To simplify the notation we use a variation of the Einstein summation convention when we take derivatives with respect to the matrix elements of $P$. For instance we have
	\begin{equation}
		\label{eq:Einstein1}
		\pdv{f}{P_a}\pdv{P_a}{g} \overset{\rm def}{=} \sum_{i,j}\pdv{f}{[P_{11}]_{i,j}} \pdv{[P_{11}]_{i,j}}{g} + \frac{1}{2}\sum_{i,j}\pdv{f}{[P_{12}]_{i,j}} \pdv{[P_{12}]_{i,j}}{g}+
		\frac{1}{2}\sum_{i,j}\pdv{f}{[P_{21}]_{i,j}} \pdv{[P_{21}]_{i,j}}{g}\,.
 	\end{equation} 
	The factor $1/2$ in the last two terms takes into account the fact that the derivative with respect to $[P_{12}]_{i,j}$ appears twice in the summation since $[P_{12}]_{i,j} = -[P_{12}]_{j,i}$ (see~\eqref{eq:P_block} and~\eqref{eq:expval_P_relations}). A usual, a complex number $z = x+iy$ and its conjugate $z^* = x-iy$ are treated as independent variables (specifically $[P_{11}]_{i,j}$ and $[P_{11}]^*_{i,j}= [P_{11}]_{j,i}$; $[P_{12}]_{i,j}$ and $[P_{21}]_{j,i} = [P_{12}]^*_{i,j}$) since $\pdv{\Omega}{z}\pdv{z}{A_l} + \pdv{\Omega}{z^*}\pdv{z^*}{A_l} = \pdv{\Omega}{x}\pdv{x}{A_l}+\pdv{\Omega}{y}\pdv{y}{A_l}$. Therefore, it is clear that the derivative of each independent matrix element of $P$ appears only once in~\eqref{eq:Einstein1}. Different subscripts $a,b,c,\dots$ added to $P$ as in the definition~\eqref{eq:Einstein1} are used to keep track of multiple partial derivatives with respect to the matrix elements of the one-particle density matrix.
	
	The first full derivative of the mean-field grand potential is simply the average current density
	\begin{equation}
		\label{eq:current_mf}
		\begin{split}
		\dv{\Omega_{\rm m.f.}\big(\vb{A},P(\vb{A}))}{A_l} &= \pdv{\Omega_{\rm m.f.}}{A_l} + \pdv{\Omega_{\rm m.f.}}{P_a}\pdv{P_a}{A_l} = \pdv{\Omega_{\rm m.f.}}{A_l} \\ &= \Tr[\pdv{K(\vb{A})}{A_l}P_{11}(\vb{A})] = -\mathcal{A}\langle \hat{J}_l(\vb{q} = \vb{0})\rangle\,.
		\end{split}
	\end{equation}
	In the second equality we have used the fact that the one-particle density matrix $P(\vb{A})$ minimizes the mean-field grand potential for each value of $\vb{A}$, thus the collection of first derivatives with respect to the matrix elements of $P$ vanishes
	\begin{equation}
		\label{eq:Omega_mf_deriv_P}
		\pdv{\Omega_{\rm m.f.}(\vb{A},P)}{P_a}\bigg|_{P = P(\vb{A})} = 0\,.	
	\end{equation} 
	This equation is the same as~\eqref{eq:minimizer} but with the effective fields replaced by the one-particle density matrix.
	The third equality in~\eqref{eq:current_mf} is a consequence of the fact that in~\eqref{eq:Omega_0_bar_P_final}-\eqref{eq:Omega_LWBK} the only term that gives a direct dependence on $\vb{A}$ is $\Tr[K(\vb{A})P_{11}] = \langle \ham_{\rm free}(\vb{A})\rangle$, while the last equality comes from~\eqref{eq:h_A_current}.
	
	The result in~\eqref{eq:current_mf} is valid for arbitrary $\vb{A}$, therefore, for the second full derivative of the grand potential we have
	\begin{equation}
		\label{eq:intermediate_2}
		\begin{split}
		\frac{d^2\Omega_{\rm m.f.}\big(\vb{A},P(\vb{A}))}{dA_l dA_m} &= \dv{}{A_m} \Tr[\pdv{K(\vb{A})}{A_l}P_{11}(\vb{A})] \\ &= \Tr[\pdv{K(\vb{A})}{A_l}{A_m}P_{11}(\vb{A})] + \Tr[\pdv{K(\vb{A})}{A_l}\pdv{P_{11}(\vb{A})}{A_m}]\,.
		\end{split}
	\end{equation}
	To proceed we need to express the partial derivative $\pdv*{P(\vb{A})}{A_m}$ in terms of quantities evaluated at $\vb{A} = \vb{0}$. To this end, we use the method proposed in Ref.~\onlinecite{Huhtinen2022} and take the full derivative of both sides of~\eqref{eq:Omega_mf_deriv_P} (compare with the discussion around~\eqref{eq:minimizer})
	\begin{equation}
		\label{eq:intermediate_1}
		0 = \dv{}{A_m}\pdv{\Omega_{\rm m.f.}(\vb{A},P(\vb{A}))}{P_a} = \pdv{\Omega_{\rm m.f.}}{A_m}{P_a} + \pdv{\Omega_{\rm m.f.}}{P_b}{P_a}\pdv{P_b}{A_m}\,.
	\end{equation}
	The first term on the right hand side is easy to evaluate since
	\begin{gather}
		\pdv{\Omega_{\rm m.f.}}{A_m}{[P_{11}]_{j,i}} = \pdv{}{[P_{11}]_{j,i}}\Tr[\pdv{K(\vb{A})}{A_m}P_{11}] = \pdv{K_{i,j}(\vb{A})}{A_m}\,,\\
		\pdv{\Omega_{\rm m.f.}}{A_m}{[P_{12}]_{j,i}} = \pdv{\Omega_{\rm m.f.}}{A_m}{[P_{21}]_{j,i}} = 0\,.
	\end{gather}
	On the other hand, we have from~\eqref{eq:Omega_LWBK}
	\begin{equation}
			\label{eq:omega_mf_omega_0_V}
			\pdv{\Omega_{\rm m.f.}}{P_b}{P_a} = \pdv{\overline{\Omega}_0}{P_b}{P_a} + \pdv{V}{P_b}{P_a}\,.
	\end{equation}
	It is straightforward to evaluate the second term on the right from the definition of the quadratic form $V(P)$. Because of~\eqref{eq:omega_deriv_inv}, the first term is identified with the Jacobian matrix of the mapping $ \expval*{\hat{c}_{i}^\dagger\hat{c}_{j}},\expval*{\hat{c}_{i}^\dagger \hat{c}_{j}^\dagger},\,\expval*{\hat{c}_{j} \hat{c}_{i}} \to \Gamma_{i,j},\,\Delta_{i,j},\,\Delta_{i,j}^*$ which we denote as
	\begin{equation}
		\label{eq:jacobian_1}
		\pdv{\Delta^a}{P_b} \overset{\rm def}{=} -\pdv{\overline{\Omega}_0}{P_b}{P_a} =
		\mqty(\pdv{\Gamma_{i',j'}}{\expval*{\hat{c}_i^\dg\hat{c}_j}} & \pdv{\Gamma_{i',j'}}{\expval*{\hat{c}_j\hat{c}_i}} & 
		\pdv{\Gamma_{i',j'}}{\expval*{\hat{c}_i^\dg\hat{c}_j^\dg}}\\ 
		\pdv{\Delta_{i',j'}}{\expval*{\hat{c}_i^\dg\hat{c}_j}} & \pdv{\Delta_{i',j'}}{\expval*{\hat{c}_j\hat{c}_i}} & 
		\pdv{\Delta_{i',j'}}{\expval*{\hat{c}_i^\dg\hat{c}_j^\dg}} \\ \pdv{\Delta^*_{i',j'}}{\expval*{\hat{c}_i^\dg\hat{c}_j}} & \pdv{\Delta^*_{i',j'}}{\expval*{\hat{c}_j\hat{c}_i}} & 
		\pdv{\Delta^*_{i',j'}}{\expval*{\hat{c}_i^\dg\hat{c}_j^\dg}})\,,
	\end{equation}
	where in the matrix on the right hand side each entry stands for the collection of derivatives labelled by all possible site indices.	
	Here and in the following, we denote by $\Delta^a = (\Gamma_{i,j}, \Delta_{i,j}, \Delta^*_{i,j})$ the collection of all independent effective fields, and we use a similar notation as in~\eqref{eq:Einstein1}
	\begin{equation}
		\label{eq:Einstein2}
		\pdv{f}{\Delta^a}\pdv{\Delta^a}{g} \overset{\rm def}{=} \sum_{i,j}\pdv{f}{\Gamma_{i,j}} \pdv{\Gamma_{i,j}}{g} + \frac{1}{2}\sum_{i,j}\pdv{f}{\Delta_{i,j}} \pdv{\Delta_{i,j}}{g}+
		\frac{1}{2}\sum_{i,j}\pdv{f}{\Delta^*_{i,j}} \pdv{\Delta^*_{i,j}}{g}\,.
	\end{equation}  
	Since the inverse mapping $\Gamma_{i,j},\,\Delta_{i,j},\,\Delta_{i,j}^* \to \expval*{\hat{c}_{i}^\dg\hat{c}_{j}},\,\expval*{\hat{c}_{i}^\dg \hat{c}_{j}^\dg},\,\expval*{\hat{c}_{j}\hat{c}_{i}}$ exists, we denote its Jacobian, the inverse of the matrix in~\eqref{eq:jacobian_1}, as $\pdv*{P_a}{\Delta^b}$ and we have in our notation
	\begin{equation}
			\pdv{P_a}{\Delta^b}\pdv{\Delta^b}{P_c} = \pdv{\Delta^c}{P_b}\pdv{P_b}{\Delta^a} = \delta^c_{a}\,,
	\end{equation}
	where $\delta_{a}^c$ is the Kronecker delta.
	It is also clear from~\eqref{eq:omega_deriv_1}-\eqref{eq:omega_deriv_2} that
	\begin{equation}
		\label{eq:omega_papb_dadb}
		-\qty[\pdv{\overline{\Omega}_0}{P_b}{P_a}]^{-1} = \pdv{P_a}{\Delta^b} = \pdv{\Omega_0}{\Delta^b}{\Delta^a}\,.
	\end{equation}
	Note that here we denote with $\qty[\pdv*{\overline{\Omega}_0}{P_b}{P_a}]^{-1}$ the inverse of $\pdv*{\overline{\Omega}_0}{P_b}{P_a}$ as a matrix and not the inverse of a single matrix element.   
	The relation in~\eqref{eq:omega_papb_dadb} is useful since $\pdv*{\Omega_0}{\Delta^b}{\Delta^a}$ is a collection of correlation functions similar to the one found in~\eqref{eq:chi_p} with the crucial difference that they are evaluated on the statistical ensemble given by the mean-field Hamiltonian $\ham_{\rm m.f.}$.
	These objects are easy to compute and their explicit expression is provided in Appendix~\ref{app:equivalence}.
	
	From~\eqref{eq:intermediate_1},~\eqref{eq:omega_mf_omega_0_V} and~\eqref{eq:omega_papb_dadb}, we can now express the derivative of the density matrix as
	\begin{equation}
		\pdv{P_b}{A_m} = \qty[-\qty[\pdv{\Omega_0}{\Delta^b}{\Delta^a}]^{-1}+\pdv{V}{P_b}{P_a}]^{-1}J^a_m\,,\qq{with} J^a_m \overset{\rm def}{=} - \pdv{\Omega_{\rm m.f.}}{A_m}{P_a}\,.
	\end{equation}
	Inserting this into~\eqref{eq:intermediate_2} gives our main result for the superfluid weight 
	\begin{equation}
		\label{eq:main_Ds_GRPA}
		\begin{split}
		D_{{\rm s},lm} &\approx \frac{1}{\mathcal{A}}\frac{d^2\Omega_{\rm m.f.}\big(\vb{A},P(\vb{A}))}{dA_l dA_m}\bigg|_{\vb{A}=\vb{0}} = \frac{1}{\mathcal{A}} \expval{ \pdv{\ham_{\rm free}(\vb{A})}{A_l}{A_m}\bigg|_{\vb{A}=\vb{0}}} - \frac{1}{\mathcal{A}}J_l^b\pdv{P_b}{A_m} \\
		&= -\chi_{lm}^{\rm d}(\vb{q} = \vb{0},\vb{q}'=\vb{0}) +\frac{1}{\mathcal{A}}J_l^b\qty[\qty[\pdv{\Omega_0}{\Delta^b}{\Delta^a}]^{-1}-\pdv{V}{P_b}{P_a}]^{-1}J^a_m\,.
		\end{split}
	\end{equation}
	In the last equality we have identified the diamagnetic response function~\eqref{eq:chi_d}, with the only difference that here the expectation value is evaluated on the mean-field statistical ensemble. On the other hand, the remaining term
	\begin{equation}
		\label{eq:chi_grpa}
		\chi_{lm}^{\rm p,GRPA} = -\frac{1}{\mathcal{A}}J_l^b\qty[\qty[\pdv{\Omega_0}{\Delta^b}{\Delta^a}]^{-1}-\pdv{V}{P_b}{P_a}]^{-1}J^a_m
	\end{equation}
	is not simply the mean-field version of the paramagnetic response~\eqref{eq:chi_p}. Infact, the usual mean-field paramagnetic response is obtained by setting $\pdv*{V}{P_b}{P_a} = 0$ in the above result, as explained in Appendix~\ref{app:equivalence}. This corresponds to the response function computed in Ref.~\onlinecite{Scalapino1993} and~\onlinecite{Liang2017a} for instance. On the other hand, the response function $\chi_{lm}^{\rm p,GRPA}$ has the same structure as the well-known result for the density-density response function within the random phase approximation~\cite{Giuliani2005} 
	\begin{equation}
		\label{eq:chi_rpa}
		\chi^{\rm RPA}(\vb{q},\omega) = \frac{1}{\chi_0^{-1}(\vb{q},\omega)-v(\vb{q})}\,,	
	\end{equation}  
	where $\chi_0^{-1}(\vb{q},\omega)$ is the density-density response function in the absence of interactions (the Lindhard response function~\cite{Giuliani2005}) and $v(\vb{q})$ the Fourier transform of the interaction potential (translational invariance has been assumed). Indeed, the superfluid weight within the usual (not generalized) random phase approximation is obtained from~\eqref{eq:chi_grpa} by retaining only the \qql Hartree term\qqr in the quadratic form in~\eqref{eq:Omega_LWBK}, namely by taking $V(P) = \frac{1}{2}\sum_{i,j}V_{i,j}\ev{\hat{n}_i}\ev{\hat{n}_j}$. Retaining also the \qql Fock\qqr or \qql exchange term\qqr, that is if $V(P) = \frac{1}{2}\sum_{i,j} V_{i,j}\qty\big(\expval{\hat{n}_{i}}\expval{\hat{n}_{j}}-\expval*{\hat{c}_{i}^\dagger\hat{c}_{j}}\expval*{\hat{c}_{j}^\dagger\hat{c}_{i}})$,	leads to the  time-dependent Hartree-Fock approximation, which is discussed in Ref.~\onlinecite{Giuliani2005}.  The  time-dependent Hartree-Fock approximation is the next step beyond the time-dependent Hartree approximation, which is an alternative name for the standard random phase approximation~\eqref{eq:chi_rpa}. By extension, one may call~\eqref{eq:chi_grpa} the time-dependent Hartree-Fock-Bogoliubov approximation since all terms obtained from the application of Wick's theorem in~\eqref{eq:inter_Bog} are retained in $V(P)$, including the \qql pairing\qqr or \qql Bogoliubov term\qqr $\expval*{\hat{c}_{i}^\dg\hat{c}_{j}^\dg}\expval*{\hat{c}_{j}\hat{c}_{i}}$. 
	
	
	The generalization of the random phase approximation for superconductors was originally developed originally by Anderson~\cite{Anderson1958} and Rickayzen~\cite{Rickayzen1959} (see also Ref.~\onlinecite{Schrieffer1964}), however the result for the superfluid weight in the simple form of~\eqref{eq:main_Ds_GRPA} has never been presented before.  Most importantly, it was not appreciated before our work that the generalized random phase approximation can be obtained by simply replacing the exact grand potential with the mean-field one in~\eqref{eq:d2F_dA2} and taking into account the $\vb{A}$-dependence of the effective fields (see~\eqref{eq:D_mf}). We have argued that this procedure is more theoretically sound since mean-field theory is a variational approximation for the grand potential, as explained in Sec.~\ref{sec:mean-field}, and gauge invariance is preserved. Indeed, one of the main reasons for introducing the generalized random phase approximation is to cure the problem of the loss of gauge invariance occurring when the superfluid weight and other observables are evaluated by replacing the exact Hamiltonian with the mean-field one in the correlation functions~\eqref{eq:chi_d} and~\eqref{eq:chi_p}. 
	
	We expect our gauge invariant result~\eqref{eq:main_Ds_GRPA} to be useful to deal with the problems pointed out in Refs.~\onlinecite{Chan2022a} and~\onlinecite{Huhtinen2022} regarding the evaluation of the superfluid weight in multiorbital lattices. It would also be interesting to investigate whether~\eqref{eq:main_Ds_GRPA} can be implemented numerically in a way which scales linearly with the system size. This would be the case if it turns out that the matrix of correlation functions $\pdv*{\Omega_0}{\Delta^b}{\Delta^a}$, or better its inverse $-\pdv*{\overline{\Omega}_0}{P_b}{P_a}$, can be approximately represented by a sparse matrix. To this end, it is important to estimate the decay behavior with distance of the elements of these matrices, as in the case of the one-particle density matrix~\eqref{eq:density_matrix_decay}, for which little is known in the case of superconductive systems. We leave the answers to these interesting questions for the future.
	
	\section{Conclusion and perspectives}
	\label{sec:conclusion}
	
	In this work, we have begun exploring the advantages of reformulating mean-field theory for superconducting systems in terms of the one-particle density matrix. There is the promise of significant computational advantages since~\eqref{eq:Omega_0_bar_P_final}-\eqref{eq:Omega_LWBK} provide the starting point for applying to superconducting systems the linear scaling methods developed in the context of electronic structure theory~\cite{Goedecker1999,Bowler2012}. Whether this approach will allow for the simulation of large supercells of disordered superconductors, larger than it is currently possible, ultimately depends on how fast the off-diagonal matrix elements of the density matrix decay with distance. As mentioned before, further work in this direction is required in the case of superconducting systems. The application of linear scaling methods to superconductors with many atoms in the unit cell, such as the recently discovered twisted bilayer graphene and other moir\'e materials, is another important motivation for the present work. As discussed in Section~\ref{sec:mermin_functional}, it should be rather straightforward to adapt to superconducting systems the linear scaling algorithms that are currently available.
	
	By writing the mean-field grand potential as an explicit function of the density matrix, we have also elucidated the relation between two popular methods used to study superconducting systems: the mean-field approximation and the generalized random phase approximation, which is usually considered a beyond mean-field approximation. Indeed, we have calculated the superfluid weight as the full second derivative of the grand potential and found a general and gauge invariant result~\eqref{eq:main_Ds_GRPA}, which differs from the standard mean-field result used in the literature~\eqref{eq:Ds_mf} (with the exception of Refs.~\onlinecite{Huhtinen2022} and~\onlinecite{Chan2022a}). The difference is in the paramagnetic current-current response function~\eqref{eq:chi_grpa}, which has the form typical of the random phase approximation~\eqref{eq:chi_rpa}. This is an example of how formulating mean-field theory in terms of the density matrix can be advantageous for analytical calculations since our derivation of~\eqref{eq:main_Ds_GRPA} is very transparent. Our result could be useful for solving the problems encountered when evaluating the superfluid weight in multiorbital systems, in particular when flat bands are present, as pointed out in Refs.~\onlinecite{Huhtinen2022} and~\onlinecite{Chan2022a}. 
	
	From a numerical point of view it is an interesting question whether~\eqref{eq:main_Ds_GRPA} can be used to compute the superfluid weight with a computational cost that scales linearly with the system size. This depends on whether the Jacobian matrix~\eqref{eq:jacobian_1} has matrix elements that decay rapidly with distance, in the same way as the density matrix~\eqref{eq:density_matrix_decay}, another interesting and important question for the future. Finally, it would  be also important to extend some of our results to include more realistic interactions, such as the retarded electron-phonon interaction, which is responsible for the formation of Cooper pairs in many superconducting materials. To this end, it could be interesting to combine, if possible, our formulation with Eliashberg theory, used for superconductors with strong electron-phonon coupling~\cite{Ummarino2017}.
	
	At a general level we hope that our work will stimulate the adoption of density matrix-based methods in the context of superconductivity.
	
	\acknowledgments

	We thank Koushik Swaminathan, Pham Nguyen Minh, Riku Tuovinen, Ville Pyykk\"onen, Jami Kinnunen, Kukka-Emilia Huhtinen, P\"aivi T\"orm\"a and Giovanni Vignale for useful discussions and proofreading.
	This work has been supported by the Academy of Finland under Grants No. 330384
	and No. 336369.

	\appendix
	
	\section{Evaluation of~\eqref{eq:omega_papb_dadb}}
	\label{app:equivalence}
	
	Ref.~\onlinecite{Feynman1972} gives the following perturbative expansion for the partition function
	\begin{equation}\label{eq:Feynman_result}
		\Tr\qty\big[e^{-\beta(\ham+\vam)}] = \Tr\qty\big[e^{-\beta\ham}] - \beta \Tr\qty\big[e^{-\beta\ham}\vam] +\frac{\beta}{2}\int_0^\beta\dd{w} \Tr\qty\big[e^{-\beta\ham}e^{w \ham}\vam e^{-w\ham}\vam] + \dots
	\end{equation} 
	We apply this general result with $\ham$ a generic Hamiltonian and $\vam$ a generic quadratic Hamiltonian (see~\eqref{eq:H_BdG})
	\begin{equation}
	 \vam = \sum_{i,j} \qty(\Gamma_{i,j} \hat{c}_{i}^\dg \hat{c}_{j} + \frac{1}{2}\Delta_{i,j} \hat{c}_{i}^\dg \hat{c}_{j}^\dg + \frac{1}{2}\Delta_{i,j}^*\hat{c}_{j}\hat{c}_{i})\,.
	\end{equation}
	Using~\eqref{eq:Feynman_result} the partial derivatives of the grand potential $\Omega = -\beta^{-1}\ln\Tr\qty\big[e^{\beta(\ham+\vam)}]$ can be evaluated as follows
	\begin{equation}
		\label{eq:Omega_correlation}
		\pdv{\Omega}{\Gamma_{i,j}}{\Gamma_{i',j'}}\bigg|_{\Gamma,\Delta = 0} = -\int_0^{\beta}\dd{w} \expval{e^{w\ham}\qty\big(\hat{c}_i^\dg\hat{c}_j -\expval*{\hat{c}_i^\dg\hat{c}_j})e^{-w\ham}\qty\big(\hat{c}_{i'}^\dg\hat{c}_{j'} -\expval*{\hat{c}_{i'}^\dg\hat{c}_{j'}})}\,,	
	\end{equation}
	with $\langle\, \cdot\,\rangle = \Tr\big[\cdot\,e^{-\beta \ham}\big]/\mathcal{Z}$. By replacing the operators $\hat{c}_i^\dg\hat{c}_j$ with $\hat{c}_i^\dg\hat{c}_j^\dg$ and $\hat{c}_i\hat{c}_j$ in the above equation, the analogous results for the derivatives with respect to $\Delta_{i,j}$ and $\Delta_{i,j}^*$ are readily obtained.  If the Hamiltonian $\ham$  coincides with the one in~\eqref{eq:H_BdG}, we obtain the expression for the partial derivatives of $\Omega_0$ in~\eqref{eq:omega_papb_dadb} as correlation functions on the mean-field statistical ensemble. These are important for the  paramagnetic response function in the generalized random phase approximation~\eqref{eq:chi_grpa}.
	
	Neglecting the term $\pdv*{V}{P_a}{P_b}$ in~\eqref{eq:chi_grpa} and using~\eqref{eq:Omega_correlation} leads to the standard mean-field result for the superfluid weight~\cite{Scalapino1993,Liang2017a}
	\begin{equation}
		\label{eq:Ds_mf}
		D_{{\rm s},ml} = \frac{1}{\mathcal{A}} \expval{ \pdv{\ham_{\rm free}(\vb{A})}{A_l}{A_m}\bigg|_{\vb{A}=\vb{0}}}-\mathcal{A}\int_0^{\beta}\dd{w} \expval{e^{w\ham_0}\qty\big(\hat{J}_l -\expval*{\hat{J}_l})e^{-w\ham_0}\qty\big(\hat{J}_m -\expval*{\hat{J}_m})}\,,
	\end{equation}
	with $\hat{J}_l \overset{\rm def}{=} \hat{J}_l(\vb{q}=\vb{0})$. We stress that here all the expectation values are taken on the mean-field statistical ensemble.
	The correlation function that appears here is  in principle different from the one in~\eqref{eq:chi_p}, which is the retarded response function. Indeed, \eqref{eq:Ds_mf} corresponds to the so-called isothermal response to an external perturbation (for a discussion on the difference between isothermal and adiabatic responses see Ref.~\onlinecite{Giuliani2005}). However, one can check in simple cases that~\eqref{eq:superfluid_weight}, computed using the mean-field statistical ensemble, and~\eqref{eq:Ds_mf} give the  same result if superconducting order is present in the system~\cite{Scalapino1993}.


\begin{thebibliography}{47}
\expandafter\ifx\csname natexlab\endcsname\relax\def\natexlab#1{#1}\fi
\expandafter\ifx\csname bibnamefont\endcsname\relax
  \def\bibnamefont#1{#1}\fi
\expandafter\ifx\csname bibfnamefont\endcsname\relax
  \def\bibfnamefont#1{#1}\fi
\expandafter\ifx\csname citenamefont\endcsname\relax
  \def\citenamefont#1{#1}\fi
\expandafter\ifx\csname url\endcsname\relax
  \def\url#1{\texttt{#1}}\fi
\expandafter\ifx\csname urlprefix\endcsname\relax\def\urlprefix{URL }\fi
\providecommand{\bibinfo}[2]{#2}
\providecommand{\eprint}[2][]{\url{#2}}

\bibitem[{\citenamefont{Sadovskii}(1997)}]{Sadovskii1997}
\bibinfo{author}{\bibfnamefont{M.~V.} \bibnamefont{Sadovskii}},
  \bibinfo{journal}{Physics Reports} \textbf{\bibinfo{volume}{282}},
  \bibinfo{pages}{225} (\bibinfo{year}{1997}),
  \urlprefix\url{https://www.sciencedirect.com/science/article/pii/S0370157396000361}.

\bibitem[{\citenamefont{Imry and Strongin}(1981)}]{Imry1981}
\bibinfo{author}{\bibfnamefont{Y.}~\bibnamefont{Imry}} \bibnamefont{and}
  \bibinfo{author}{\bibfnamefont{M.}~\bibnamefont{Strongin}},
  \bibinfo{journal}{Physical Review B} \textbf{\bibinfo{volume}{24}},
  \bibinfo{pages}{6353} (\bibinfo{year}{1981}),
  \urlprefix\url{https://link.aps.org/doi/10.1103/PhysRevB.24.6353}.

\bibitem[{\citenamefont{Maekawa et~al.}(1984)\citenamefont{Maekawa, Ebisawa,
  and Fukuyama}}]{Maekawa1984}
\bibinfo{author}{\bibfnamefont{S.}~\bibnamefont{Maekawa}},
  \bibinfo{author}{\bibfnamefont{H.}~\bibnamefont{Ebisawa}}, \bibnamefont{and}
  \bibinfo{author}{\bibfnamefont{H.}~\bibnamefont{Fukuyama}},
  \bibinfo{journal}{Journal of the Physical Society of Japan}
  \textbf{\bibinfo{volume}{53}}, \bibinfo{pages}{2681} (\bibinfo{year}{1984}),
  \urlprefix\url{https://journals.jps.jp/doi/abs/10.1143/JPSJ.53.2681}.

\bibitem[{\citenamefont{Ma et~al.}(1986)\citenamefont{Ma, Halperin, and
  Lee}}]{Ma1986}
\bibinfo{author}{\bibfnamefont{M.}~\bibnamefont{Ma}},
  \bibinfo{author}{\bibfnamefont{B.~I.} \bibnamefont{Halperin}},
  \bibnamefont{and} \bibinfo{author}{\bibfnamefont{P.~A.} \bibnamefont{Lee}},
  \bibinfo{journal}{Physical Review B} \textbf{\bibinfo{volume}{34}},
  \bibinfo{pages}{3136} (\bibinfo{year}{1986}),
  \urlprefix\url{https://link.aps.org/doi/10.1103/PhysRevB.34.3136}.

\bibitem[{\citenamefont{Ghosal et~al.}(1998)\citenamefont{Ghosal, Randeria, and
  Trivedi}}]{Ghosal1998}
\bibinfo{author}{\bibfnamefont{A.}~\bibnamefont{Ghosal}},
  \bibinfo{author}{\bibfnamefont{M.}~\bibnamefont{Randeria}}, \bibnamefont{and}
  \bibinfo{author}{\bibfnamefont{N.}~\bibnamefont{Trivedi}},
  \bibinfo{journal}{Physical Review Letters} \textbf{\bibinfo{volume}{81}},
  \bibinfo{pages}{3940} (\bibinfo{year}{1998}),
  \urlprefix\url{https://link.aps.org/doi/10.1103/PhysRevLett.81.3940}.

\bibitem[{\citenamefont{Ghosal et~al.}(2001)\citenamefont{Ghosal, Randeria, and
  Trivedi}}]{Ghosal2001}
\bibinfo{author}{\bibfnamefont{A.}~\bibnamefont{Ghosal}},
  \bibinfo{author}{\bibfnamefont{M.}~\bibnamefont{Randeria}}, \bibnamefont{and}
  \bibinfo{author}{\bibfnamefont{N.}~\bibnamefont{Trivedi}},
  \bibinfo{journal}{Physical Review B} \textbf{\bibinfo{volume}{65}},
  \bibinfo{pages}{014501} (\bibinfo{year}{2001}),
  \urlprefix\url{https://link.aps.org/doi/10.1103/PhysRevB.65.014501}.

\bibitem[{\citenamefont{Zhao et~al.}(2019)\citenamefont{Zhao, Lin, Xiao, Huang,
  Yao, Yan, Xing, Zhang, Li, Hoshino et~al.}}]{Zhao2019}
\bibinfo{author}{\bibfnamefont{K.}~\bibnamefont{Zhao}},
  \bibinfo{author}{\bibfnamefont{H.}~\bibnamefont{Lin}},
  \bibinfo{author}{\bibfnamefont{X.}~\bibnamefont{Xiao}},
  \bibinfo{author}{\bibfnamefont{W.}~\bibnamefont{Huang}},
  \bibinfo{author}{\bibfnamefont{W.}~\bibnamefont{Yao}},
  \bibinfo{author}{\bibfnamefont{M.}~\bibnamefont{Yan}},
  \bibinfo{author}{\bibfnamefont{Y.}~\bibnamefont{Xing}},
  \bibinfo{author}{\bibfnamefont{Q.}~\bibnamefont{Zhang}},
  \bibinfo{author}{\bibfnamefont{Z.-X.} \bibnamefont{Li}},
  \bibinfo{author}{\bibfnamefont{S.}~\bibnamefont{Hoshino}},
  \bibnamefont{et~al.}, \bibinfo{journal}{Nature Physics}
  \textbf{\bibinfo{volume}{15}}, \bibinfo{pages}{904} (\bibinfo{year}{2019}),
  \urlprefix\url{https://www.nature.com/articles/s41567-019-0570-0}.

\bibitem[{\citenamefont{Sac{\'e}p{\'e}
  et~al.}(2020)\citenamefont{Sac{\'e}p{\'e}, Feigel'man, and
  Klapwijk}}]{Sacepe2020}
\bibinfo{author}{\bibfnamefont{B.}~\bibnamefont{Sac{\'e}p{\'e}}},
  \bibinfo{author}{\bibfnamefont{M.}~\bibnamefont{Feigel'man}},
  \bibnamefont{and} \bibinfo{author}{\bibfnamefont{T.~M.}
  \bibnamefont{Klapwijk}}, \bibinfo{journal}{Nature Physics}
  \textbf{\bibinfo{volume}{16}}, \bibinfo{pages}{734} (\bibinfo{year}{2020}),
  \urlprefix\url{https://www.nature.com/articles/s41567-020-0905-x}.

\bibitem[{\citenamefont{Lau et~al.}(2022)\citenamefont{Lau, Peotta, Pikulin,
  Rossi, and Hyart}}]{Lau2022}
\bibinfo{author}{\bibfnamefont{A.}~\bibnamefont{Lau}},
  \bibinfo{author}{\bibfnamefont{S.}~\bibnamefont{Peotta}},
  \bibinfo{author}{\bibfnamefont{D.~I.} \bibnamefont{Pikulin}},
  \bibinfo{author}{\bibfnamefont{E.}~\bibnamefont{Rossi}}, \bibnamefont{and}
  \bibinfo{author}{\bibfnamefont{T.}~\bibnamefont{Hyart}},
  \bibinfo{journal}{arXiv:2203.01058 [cond-mat]}  (\bibinfo{year}{2022}),
  \eprint{2203.01058}, \urlprefix\url{http://arxiv.org/abs/2203.01058}.

\bibitem[{\citenamefont{London et~al.}(1935)\citenamefont{London, London, and
  Lindemann}}]{London1935}
\bibinfo{author}{\bibfnamefont{F.}~\bibnamefont{London}},
  \bibinfo{author}{\bibfnamefont{H.}~\bibnamefont{London}}, \bibnamefont{and}
  \bibinfo{author}{\bibfnamefont{F.~A.} \bibnamefont{Lindemann}},
  \bibinfo{journal}{Proceedings of the Royal Society of London. Series A -
  Mathematical and Physical Sciences} \textbf{\bibinfo{volume}{149}},
  \bibinfo{pages}{71} (\bibinfo{year}{1935}),
  \urlprefix\url{https://royalsocietypublishing.org/doi/10.1098/rspa.1935.0048}.

\bibitem[{\citenamefont{Scalapino et~al.}(1992)\citenamefont{Scalapino, White,
  and Zhang}}]{Scalapino1992}
\bibinfo{author}{\bibfnamefont{D.~J.} \bibnamefont{Scalapino}},
  \bibinfo{author}{\bibfnamefont{S.~R.} \bibnamefont{White}}, \bibnamefont{and}
  \bibinfo{author}{\bibfnamefont{S.~C.} \bibnamefont{Zhang}},
  \bibinfo{journal}{Physical Review Letters} \textbf{\bibinfo{volume}{68}},
  \bibinfo{pages}{2830} (\bibinfo{year}{1992}),
  \urlprefix\url{https://link.aps.org/doi/10.1103/PhysRevLett.68.2830}.

\bibitem[{\citenamefont{Scalapino et~al.}(1993)\citenamefont{Scalapino, White,
  and Zhang}}]{Scalapino1993}
\bibinfo{author}{\bibfnamefont{D.~J.} \bibnamefont{Scalapino}},
  \bibinfo{author}{\bibfnamefont{S.~R.} \bibnamefont{White}}, \bibnamefont{and}
  \bibinfo{author}{\bibfnamefont{S.}~\bibnamefont{Zhang}},
  \bibinfo{journal}{Physical Review B} \textbf{\bibinfo{volume}{47}},
  \bibinfo{pages}{7995} (\bibinfo{year}{1993}),
  \urlprefix\url{https://link.aps.org/doi/10.1103/PhysRevB.47.7995}.

\bibitem[{\citenamefont{Nelson and Kosterlitz}(1977)}]{Nelson1977}
\bibinfo{author}{\bibfnamefont{D.~R.} \bibnamefont{Nelson}} \bibnamefont{and}
  \bibinfo{author}{\bibfnamefont{J.~M.} \bibnamefont{Kosterlitz}},
  \bibinfo{journal}{Physical Review Letters} \textbf{\bibinfo{volume}{39}},
  \bibinfo{pages}{1201} (\bibinfo{year}{1977}),
  \urlprefix\url{https://link.aps.org/doi/10.1103/PhysRevLett.39.1201}.

\bibitem[{\citenamefont{Fisher et~al.}(1973)\citenamefont{Fisher, Barber, and
  Jasnow}}]{Fisher1973}
\bibinfo{author}{\bibfnamefont{M.~E.} \bibnamefont{Fisher}},
  \bibinfo{author}{\bibfnamefont{M.~N.} \bibnamefont{Barber}},
  \bibnamefont{and} \bibinfo{author}{\bibfnamefont{D.}~\bibnamefont{Jasnow}},
  \bibinfo{journal}{Physical Review A} \textbf{\bibinfo{volume}{8}},
  \bibinfo{pages}{1111} (\bibinfo{year}{1973}),
  \urlprefix\url{https://link.aps.org/doi/10.1103/PhysRevA.8.1111}.

\bibitem[{\citenamefont{Tovmasyan et~al.}(2018)\citenamefont{Tovmasyan, Peotta,
  Liang, T{\"o}rm{\"a}, and Huber}}]{Tovmasyan2018}
\bibinfo{author}{\bibfnamefont{M.}~\bibnamefont{Tovmasyan}},
  \bibinfo{author}{\bibfnamefont{S.}~\bibnamefont{Peotta}},
  \bibinfo{author}{\bibfnamefont{L.}~\bibnamefont{Liang}},
  \bibinfo{author}{\bibfnamefont{P.}~\bibnamefont{T{\"o}rm{\"a}}},
  \bibnamefont{and} \bibinfo{author}{\bibfnamefont{S.~D.} \bibnamefont{Huber}},
  \bibinfo{journal}{Physical Review B} \textbf{\bibinfo{volume}{98}},
  \bibinfo{pages}{134513} (\bibinfo{year}{2018}),
  \urlprefix\url{https://link.aps.org/doi/10.1103/PhysRevB.98.134513}.

\bibitem[{\citenamefont{Bardeen
  et~al.}(1957{\natexlab{a}})\citenamefont{Bardeen, Cooper, and
  Schrieffer}}]{Bardeen1957}
\bibinfo{author}{\bibfnamefont{J.}~\bibnamefont{Bardeen}},
  \bibinfo{author}{\bibfnamefont{L.~N.} \bibnamefont{Cooper}},
  \bibnamefont{and} \bibinfo{author}{\bibfnamefont{J.~R.}
  \bibnamefont{Schrieffer}}, \bibinfo{journal}{Physical Review}
  \textbf{\bibinfo{volume}{106}}, \bibinfo{pages}{162}
  (\bibinfo{year}{1957}{\natexlab{a}}),
  \urlprefix\url{https://link.aps.org/doi/10.1103/PhysRev.106.162}.

\bibitem[{\citenamefont{Bardeen
  et~al.}(1957{\natexlab{b}})\citenamefont{Bardeen, Cooper, and
  Schrieffer}}]{Bardeen1957a}
\bibinfo{author}{\bibfnamefont{J.}~\bibnamefont{Bardeen}},
  \bibinfo{author}{\bibfnamefont{L.~N.} \bibnamefont{Cooper}},
  \bibnamefont{and} \bibinfo{author}{\bibfnamefont{J.~R.}
  \bibnamefont{Schrieffer}}, \bibinfo{journal}{Physical Review}
  \textbf{\bibinfo{volume}{108}}, \bibinfo{pages}{1175}
  (\bibinfo{year}{1957}{\natexlab{b}}),
  \urlprefix\url{https://link.aps.org/doi/10.1103/PhysRev.108.1175}.

\bibitem[{\citenamefont{Schrieffer}(1964)}]{Schrieffer1964}
\bibinfo{author}{\bibfnamefont{J.~R.} \bibnamefont{Schrieffer}},
  \emph{\bibinfo{title}{Theory of Superconductivity}}, Frontiers in Physics
  (\bibinfo{publisher}{{Benjamin}}, \bibinfo{address}{{New York}},
  \bibinfo{year}{1964}).

\bibitem[{\citenamefont{Tinkham}(2004)}]{Tinkham2004}
\bibinfo{author}{\bibfnamefont{M.}~\bibnamefont{Tinkham}},
  \emph{\bibinfo{title}{Introduction to Superconductivity}}
  (\bibinfo{publisher}{{Dover Publications}}, \bibinfo{address}{{Mineola,
  N.Y}}, \bibinfo{year}{2004}), \bibinfo{edition}{2nd} ed.

\bibitem[{\citenamefont{Liang et~al.}(2017)\citenamefont{Liang, Vanhala,
  Peotta, Siro, Harju, and T{\"o}rm{\"a}}}]{Liang2017a}
\bibinfo{author}{\bibfnamefont{L.}~\bibnamefont{Liang}},
  \bibinfo{author}{\bibfnamefont{T.~I.} \bibnamefont{Vanhala}},
  \bibinfo{author}{\bibfnamefont{S.}~\bibnamefont{Peotta}},
  \bibinfo{author}{\bibfnamefont{T.}~\bibnamefont{Siro}},
  \bibinfo{author}{\bibfnamefont{A.}~\bibnamefont{Harju}}, \bibnamefont{and}
  \bibinfo{author}{\bibfnamefont{P.}~\bibnamefont{T{\"o}rm{\"a}}},
  \bibinfo{journal}{Physical Review B} \textbf{\bibinfo{volume}{95}},
  \bibinfo{pages}{024515} (\bibinfo{year}{2017}),
  \urlprefix\url{https://link.aps.org/doi/10.1103/PhysRevB.95.024515}.

\bibitem[{\citenamefont{Huhtinen et~al.}(2022)\citenamefont{Huhtinen,
  {Herzog-Arbeitman}, Chew, Bernevig, and T{\"o}rm{\"a}}}]{Huhtinen2022}
\bibinfo{author}{\bibfnamefont{K.-E.} \bibnamefont{Huhtinen}},
  \bibinfo{author}{\bibfnamefont{J.}~\bibnamefont{{Herzog-Arbeitman}}},
  \bibinfo{author}{\bibfnamefont{A.}~\bibnamefont{Chew}},
  \bibinfo{author}{\bibfnamefont{B.~A.} \bibnamefont{Bernevig}},
  \bibnamefont{and}
  \bibinfo{author}{\bibfnamefont{P.}~\bibnamefont{T{\"o}rm{\"a}}},
  \bibinfo{journal}{Physical Review B} \textbf{\bibinfo{volume}{106}},
  \bibinfo{pages}{014518} (\bibinfo{year}{2022}),
  \urlprefix\url{https://link.aps.org/doi/10.1103/PhysRevB.106.014518}.

\bibitem[{\citenamefont{Chan et~al.}(2022)\citenamefont{Chan, Gr{\'e}maud, and
  Batrouni}}]{Chan2022a}
\bibinfo{author}{\bibfnamefont{S.~M.} \bibnamefont{Chan}},
  \bibinfo{author}{\bibfnamefont{B.}~\bibnamefont{Gr{\'e}maud}},
  \bibnamefont{and} \bibinfo{author}{\bibfnamefont{G.~G.}
  \bibnamefont{Batrouni}}, \bibinfo{journal}{Physical Review B}
  \textbf{\bibinfo{volume}{105}}, \bibinfo{pages}{024502}
  (\bibinfo{year}{2022}),
  \urlprefix\url{https://link.aps.org/doi/10.1103/PhysRevB.105.024502}.

\bibitem[{\citenamefont{Peotta and T{\"o}rm{\"a}}(2015)}]{Peotta2015}
\bibinfo{author}{\bibfnamefont{S.}~\bibnamefont{Peotta}} \bibnamefont{and}
  \bibinfo{author}{\bibfnamefont{P.}~\bibnamefont{T{\"o}rm{\"a}}},
  \bibinfo{journal}{Nature Communications} \textbf{\bibinfo{volume}{6}},
  \bibinfo{pages}{8944} (\bibinfo{year}{2015}),
  \urlprefix\url{https://www.nature.com/articles/ncomms9944}.

\bibitem[{\citenamefont{Anderson}(1958)}]{Anderson1958}
\bibinfo{author}{\bibfnamefont{P.~W.} \bibnamefont{Anderson}},
  \bibinfo{journal}{Physical Review} \textbf{\bibinfo{volume}{112}},
  \bibinfo{pages}{1900} (\bibinfo{year}{1958}),
  \urlprefix\url{https://link.aps.org/doi/10.1103/PhysRev.112.1900}.

\bibitem[{\citenamefont{Rickayzen}(1959)}]{Rickayzen1959}
\bibinfo{author}{\bibfnamefont{G.}~\bibnamefont{Rickayzen}},
  \bibinfo{journal}{Physical Review} \textbf{\bibinfo{volume}{115}},
  \bibinfo{pages}{795} (\bibinfo{year}{1959}),
  \urlprefix\url{https://link.aps.org/doi/10.1103/PhysRev.115.795}.

\bibitem[{\citenamefont{Parravicini and Grosso}(2013)}]{Parravicini2013}
\bibinfo{author}{\bibfnamefont{G.~P.} \bibnamefont{Parravicini}}
  \bibnamefont{and} \bibinfo{author}{\bibfnamefont{G.}~\bibnamefont{Grosso}},
  \emph{\bibinfo{title}{Solid State Physics}} (\bibinfo{publisher}{{Academic
  Press}}, \bibinfo{year}{2013}).

\bibitem[{\citenamefont{Mermin}(1965)}]{Mermin1965}
\bibinfo{author}{\bibfnamefont{N.~D.} \bibnamefont{Mermin}},
  \bibinfo{journal}{Physical Review} \textbf{\bibinfo{volume}{137}},
  \bibinfo{pages}{A1441} (\bibinfo{year}{1965}),
  \urlprefix\url{https://link.aps.org/doi/10.1103/PhysRev.137.A1441}.

\bibitem[{\citenamefont{Cao et~al.}(2018)\citenamefont{Cao, Fatemi, Fang,
  Watanabe, Taniguchi, Kaxiras, and {Jarillo-Herrero}}}]{Cao2018}
\bibinfo{author}{\bibfnamefont{Y.}~\bibnamefont{Cao}},
  \bibinfo{author}{\bibfnamefont{V.}~\bibnamefont{Fatemi}},
  \bibinfo{author}{\bibfnamefont{S.}~\bibnamefont{Fang}},
  \bibinfo{author}{\bibfnamefont{K.}~\bibnamefont{Watanabe}},
  \bibinfo{author}{\bibfnamefont{T.}~\bibnamefont{Taniguchi}},
  \bibinfo{author}{\bibfnamefont{E.}~\bibnamefont{Kaxiras}}, \bibnamefont{and}
  \bibinfo{author}{\bibfnamefont{P.}~\bibnamefont{{Jarillo-Herrero}}},
  \bibinfo{journal}{Nature} \textbf{\bibinfo{volume}{556}}, \bibinfo{pages}{43}
  (\bibinfo{year}{2018}),
  \urlprefix\url{https://www.nature.com/articles/nature26160}.

\bibitem[{\citenamefont{T{\"o}rm{\"a} et~al.}(2022)\citenamefont{T{\"o}rm{\"a},
  Peotta, and Bernevig}}]{Torma2022}
\bibinfo{author}{\bibfnamefont{P.}~\bibnamefont{T{\"o}rm{\"a}}},
  \bibinfo{author}{\bibfnamefont{S.}~\bibnamefont{Peotta}}, \bibnamefont{and}
  \bibinfo{author}{\bibfnamefont{B.~A.} \bibnamefont{Bernevig}},
  \bibinfo{journal}{Nature Reviews Physics} \textbf{\bibinfo{volume}{4}},
  \bibinfo{pages}{528} (\bibinfo{year}{2022}),
  \urlprefix\url{https://www.nature.com/articles/s42254-022-00466-y}.

\bibitem[{\citenamefont{Hu et~al.}(2019)\citenamefont{Hu, Hyart, Pikulin, and
  Rossi}}]{Hu2019}
\bibinfo{author}{\bibfnamefont{X.}~\bibnamefont{Hu}},
  \bibinfo{author}{\bibfnamefont{T.}~\bibnamefont{Hyart}},
  \bibinfo{author}{\bibfnamefont{D.~I.} \bibnamefont{Pikulin}},
  \bibnamefont{and} \bibinfo{author}{\bibfnamefont{E.}~\bibnamefont{Rossi}},
  \bibinfo{journal}{Physical Review Letters} \textbf{\bibinfo{volume}{123}},
  \bibinfo{pages}{237002} (\bibinfo{year}{2019}),
  \urlprefix\url{https://link.aps.org/doi/10.1103/PhysRevLett.123.237002}.

\bibitem[{\citenamefont{Julku et~al.}(2020)\citenamefont{Julku, Peltonen,
  Liang, Heikkil{\"a}, and T{\"o}rm{\"a}}}]{Julku2020}
\bibinfo{author}{\bibfnamefont{A.}~\bibnamefont{Julku}},
  \bibinfo{author}{\bibfnamefont{T.~J.} \bibnamefont{Peltonen}},
  \bibinfo{author}{\bibfnamefont{L.}~\bibnamefont{Liang}},
  \bibinfo{author}{\bibfnamefont{T.~T.} \bibnamefont{Heikkil{\"a}}},
  \bibnamefont{and}
  \bibinfo{author}{\bibfnamefont{P.}~\bibnamefont{T{\"o}rm{\"a}}},
  \bibinfo{journal}{Physical Review B} \textbf{\bibinfo{volume}{101}},
  \bibinfo{pages}{060505} (\bibinfo{year}{2020}),
  \urlprefix\url{https://link.aps.org/doi/10.1103/PhysRevB.101.060505}.

\bibitem[{\citenamefont{Xie et~al.}(2020)\citenamefont{Xie, Song, Lian, and
  Bernevig}}]{Xie2020}
\bibinfo{author}{\bibfnamefont{F.}~\bibnamefont{Xie}},
  \bibinfo{author}{\bibfnamefont{Z.}~\bibnamefont{Song}},
  \bibinfo{author}{\bibfnamefont{B.}~\bibnamefont{Lian}}, \bibnamefont{and}
  \bibinfo{author}{\bibfnamefont{B.~A.} \bibnamefont{Bernevig}},
  \bibinfo{journal}{Physical Review Letters} \textbf{\bibinfo{volume}{124}},
  \bibinfo{pages}{167002} (\bibinfo{year}{2020}),
  \urlprefix\url{https://link.aps.org/doi/10.1103/PhysRevLett.124.167002}.

\bibitem[{\citenamefont{Goedecker}(1999)}]{Goedecker1999}
\bibinfo{author}{\bibfnamefont{S.}~\bibnamefont{Goedecker}},
  \bibinfo{journal}{Reviews of Modern Physics} \textbf{\bibinfo{volume}{71}},
  \bibinfo{pages}{1085} (\bibinfo{year}{1999}),
  \urlprefix\url{https://link.aps.org/doi/10.1103/RevModPhys.71.1085}.

\bibitem[{\citenamefont{Bowler and Miyazaki}(2012)}]{Bowler2012}
\bibinfo{author}{\bibfnamefont{D.~R.} \bibnamefont{Bowler}} \bibnamefont{and}
  \bibinfo{author}{\bibfnamefont{T.}~\bibnamefont{Miyazaki}},
  \bibinfo{journal}{Reports on Progress in Physics}
  \textbf{\bibinfo{volume}{75}}, \bibinfo{pages}{036503}
  (\bibinfo{year}{2012}),
  \urlprefix\url{https://doi.org/10.1088/0034-4885/75/3/036503}.

\bibitem[{\citenamefont{Feynman}(1972)}]{Feynman1972}
\bibinfo{author}{\bibfnamefont{R.~P.} \bibnamefont{Feynman}},
  \emph{\bibinfo{title}{Statistical Mechanics: A Set of Lectures}}
  (\bibinfo{publisher}{{Reading MA}}, \bibinfo{address}{{Reading (MA)}},
  \bibinfo{year}{1972}).

\bibitem[{\citenamefont{Kuzemsky}(2015)}]{Kuzemsky2015}
\bibinfo{author}{\bibfnamefont{A.~L.} \bibnamefont{Kuzemsky}},
  \bibinfo{journal}{International Journal of Modern Physics B}
  \textbf{\bibinfo{volume}{29}}, \bibinfo{pages}{1530010}
  (\bibinfo{year}{2015}),
  \urlprefix\url{https://www.worldscientific.com/doi/abs/10.1142/S0217979215300108}.

\bibitem[{\citenamefont{{J. J. Binney} et~al.}(1992)\citenamefont{{J. J.
  Binney}, {N. J. Dowrick}, {A. J. Fisher}, and {M. E. J.
  Newman}}}]{J.J.Binney1992}
\bibinfo{author}{\bibnamefont{{J. J. Binney}}},
  \bibinfo{author}{\bibnamefont{{N. J. Dowrick}}},
  \bibinfo{author}{\bibnamefont{{A. J. Fisher}}}, \bibnamefont{and}
  \bibinfo{author}{\bibnamefont{{M. E. J. Newman}}}, \emph{\bibinfo{title}{The
  Theory of Critical Phenomena - {{An Introduction}} to the {{Renormalization
  Group}}}} (\bibinfo{publisher}{{Claredon Press}}, \bibinfo{address}{{Oxford,
  U.K}}, \bibinfo{year}{1992}).

\bibitem[{\citenamefont{Altland and Zirnbauer}(1997)}]{Altland1997}
\bibinfo{author}{\bibfnamefont{A.}~\bibnamefont{Altland}} \bibnamefont{and}
  \bibinfo{author}{\bibfnamefont{M.~R.} \bibnamefont{Zirnbauer}},
  \bibinfo{journal}{Physical Review B} \textbf{\bibinfo{volume}{55}},
  \bibinfo{pages}{1142} (\bibinfo{year}{1997}),
  \urlprefix\url{https://link.aps.org/doi/10.1103/PhysRevB.55.1142}.

\bibitem[{\citenamefont{{de Gennes}}(1966)}]{deGennes1966}
\bibinfo{author}{\bibfnamefont{P.-G.} \bibnamefont{{de Gennes}}},
  \emph{\bibinfo{title}{Superconductivity {{Of Metals And Alloys}}}}
  (\bibinfo{publisher}{{Westview Press}}, \bibinfo{year}{1966}).

\bibitem[{\citenamefont{Fetter}(2003)}]{Fetter2003}
\bibinfo{author}{\bibfnamefont{A.~L.} \bibnamefont{Fetter}},
  \emph{\bibinfo{title}{Quantum Theory of Many-Particle Systems}}
  (\bibinfo{publisher}{{Dover Publ}}, \bibinfo{address}{{Mineola, N.Y}},
  \bibinfo{year}{2003}).

\bibitem[{\citenamefont{Schnyder et~al.}(2008)\citenamefont{Schnyder, Ryu,
  Furusaki, and Ludwig}}]{Schnyder2008}
\bibinfo{author}{\bibfnamefont{A.~P.} \bibnamefont{Schnyder}},
  \bibinfo{author}{\bibfnamefont{S.}~\bibnamefont{Ryu}},
  \bibinfo{author}{\bibfnamefont{A.}~\bibnamefont{Furusaki}}, \bibnamefont{and}
  \bibinfo{author}{\bibfnamefont{A.~W.~W.} \bibnamefont{Ludwig}},
  \bibinfo{journal}{Physical Review B} \textbf{\bibinfo{volume}{78}},
  \bibinfo{pages}{195125} (\bibinfo{year}{2008}),
  \urlprefix\url{https://link.aps.org/doi/10.1103/PhysRevB.78.195125}.

\bibitem[{\citenamefont{Chiu et~al.}(2016)\citenamefont{Chiu, Teo, Schnyder,
  and Ryu}}]{Chiu2016}
\bibinfo{author}{\bibfnamefont{C.-K.} \bibnamefont{Chiu}},
  \bibinfo{author}{\bibfnamefont{J.~C.~Y.} \bibnamefont{Teo}},
  \bibinfo{author}{\bibfnamefont{A.~P.} \bibnamefont{Schnyder}},
  \bibnamefont{and} \bibinfo{author}{\bibfnamefont{S.}~\bibnamefont{Ryu}},
  \bibinfo{journal}{Reviews of Modern Physics} \textbf{\bibinfo{volume}{88}},
  \bibinfo{pages}{035005} (\bibinfo{year}{2016}),
  \urlprefix\url{https://link.aps.org/doi/10.1103/RevModPhys.88.035005}.

\bibitem[{\citenamefont{Corkill and Ho}(1996)}]{Corkill1996}
\bibinfo{author}{\bibfnamefont{J.~L.} \bibnamefont{Corkill}} \bibnamefont{and}
  \bibinfo{author}{\bibfnamefont{K.-M.} \bibnamefont{Ho}},
  \bibinfo{journal}{Physical Review B} \textbf{\bibinfo{volume}{54}},
  \bibinfo{pages}{5340} (\bibinfo{year}{1996}),
  \urlprefix\url{https://link.aps.org/doi/10.1103/PhysRevB.54.5340}.

\bibitem[{\citenamefont{{Michael Lindsey}}(2019)}]{MichaelLindsey2019}
\bibinfo{author}{\bibnamefont{{Michael Lindsey}}}, Ph.D. thesis,
  \bibinfo{school}{University of California, Berkeley} (\bibinfo{year}{2019}).

\bibitem[{\citenamefont{Giuliani and Vignale}(2005)}]{Giuliani2005}
\bibinfo{author}{\bibfnamefont{G.}~\bibnamefont{Giuliani}} \bibnamefont{and}
  \bibinfo{author}{\bibfnamefont{G.}~\bibnamefont{Vignale}},
  \emph{\bibinfo{title}{Quantum Theory of the Electron Liquid}}
  (\bibinfo{publisher}{{Cambridge University Press}},
  \bibinfo{address}{{Cambridge}}, \bibinfo{year}{2005}).

\bibitem[{\citenamefont{Taylor et~al.}(2006)\citenamefont{Taylor, Griffin,
  Fukushima, and Ohashi}}]{Taylor2006}
\bibinfo{author}{\bibfnamefont{E.}~\bibnamefont{Taylor}},
  \bibinfo{author}{\bibfnamefont{A.}~\bibnamefont{Griffin}},
  \bibinfo{author}{\bibfnamefont{N.}~\bibnamefont{Fukushima}},
  \bibnamefont{and} \bibinfo{author}{\bibfnamefont{Y.}~\bibnamefont{Ohashi}},
  \bibinfo{journal}{Physical Review A} \textbf{\bibinfo{volume}{74}},
  \bibinfo{pages}{063626} (\bibinfo{year}{2006}),
  \urlprefix\url{https://link.aps.org/doi/10.1103/PhysRevA.74.063626}.

\bibitem[{\citenamefont{Ummarino}(2017)}]{Ummarino2017}
\bibinfo{author}{\bibfnamefont{G.~A.~C.} \bibnamefont{Ummarino}},
  \emph{\bibinfo{title}{The Physics of Correlated Insulators, Metals, and
  Superconductors: lecture notes of the Autumn School on Correlated Electrons
  2017: at Forschungszentrum Jülich, 25-29 September 2017}}
  (\bibinfo{publisher}{Forschungszentrum, Zentralbibliothek},
  \bibinfo{year}{2017}), chap. \bibinfo{chapter}{13, \textit{Eliashberg
  theory}}, \urlprefix\url{https://juser.fz-juelich.de/record/837488}.

\end{thebibliography}

	\end{document}